\documentclass[runningheads]{llncs}

\usepackage{listings}
\usepackage{amssymb}

\usepackage{syntax}

\usepackage{adjustbox}

\usepackage{makecell}

\usepackage[vlined, linesnumbered, noend]{algorithm2e}

\SetCommentSty{mycommfont}

\usepackage{todonotes}

\usepackage{mathtools}
\usepackage{bm}

\usepackage{tikz}

\usepackage{ragged2e}
\usepackage{array, multirow}

\usepackage{graphicx}
\usepackage[firstpage]{draftwatermark}

\usepackage{subcaption}

\usepackage{amssymb}% http://ctan.org/pkg/amssymb
\usepackage{pifont}% http://ctan.org/pkg/pifont
\newcommand{\xmark}{\ding{55}}%

\usepackage{hyperref}
\usepackage{pgfplots}
\pgfplotsset{compat=1.14}

\usepackage{xcolor}
\usepackage{soul}
\newcommand{\mathcolorbox}[2]{%
  \begingroup\setlength{\fboxsep}{1pt}%
  \colorbox{#1}{\texttt{\hspace*{2pt}\vphantom{Ay}$#2$\hspace*{2pt}}}
  \endgroup
}

\usepackage{longtable}
\usepackage{multirow}

\newcommand{\B}{\mathbb{B}} % The set of boolean values {true, false}
\newcommand{\Z}{\mathbb{Z}}

\newcommand{\twodots}{\mathinner {\ldotp \ldotp}}
\newcommand{\Interval}[2]{[{#1}\twodots{#2}]}
\newcommand{\Theory}[1]{T_{#1}}
\newcommand{\BoundedArrayTheory}[2]{\Theory{#1}^{#2}}

\newcommand{\BoundedArrayLength}[1]{|{#1}|}
\newcommand{\BoundedArrayRead}[2]{{#1}[{#2}]}
\newcommand{\BoundedArrayWrite}[3]{{#1}\{{#2}\leftarrow{#3}\}}
\newcommand{\ArrayQuantifierSet}[2]{Q_{#1}^{#2}}
\newcommand{\QuantifierSet}[1]{Q^{#1}}
\newcommand{\PredicateOfDatapoint}[1]{L({#1})}
\newcommand{\ArrayLengthVariablesOfPredicate}[1]{S^{#1}}
\newcommand{\ValueSet}[1]{A^{#1}}
\newcommand{\Diagrams}[2]{\textsc{Diagrams}^{#1}({#2})}

\newcommand{\VariablesOf}[1]{\mathcal{D}^{#1}}
\newcommand{\VariablesTypedOf}[2]{\mathcal{D}^{#1}_{#2}}
\newcommand{\ArraysOf}[1]{\mathcal{A}^{#1}}

\newcommand{\Interpretation}[2]{{#2}}%\newcommand{\Interpretation}[2]{\lambda{ #1}.\; {#2}}
\newcommand{\Predicate}[1]{\boldsymbol{#1}}

\newcommand{\ssubsubsection}[1]{{\noindent \bf #1.}}

\newif\iflong
\longtrue
\begin{document}

\title{Data-driven Verification of Procedural Programs with Integer Arrays}

\titlerunning{Data-driven Verification of Procedural Programs with Integer Arrays}

\iflong
\author{Ahmed Bouajjani\inst{1}%\orcidID{0000-0002-2060-3592}
\and
Wael-Amine Boutglay\inst{1,2}%\orcidID{0000-0003-2068-3729}
\and
Peter Habermehl\inst{1}%\orcidID{0000-0002-7982-0946}
}
\else
\author{Ahmed Bouajjani\inst{1}\orcidID{0000-0002-2060-3592}
\and
Wael-Amine Boutglay\inst{1,2}\orcidID{0000-0003-2068-3729}
\and
Peter Habermehl\inst{1}\orcidID{0000-0002-7982-0946}
}
\fi

\authorrunning{A. Bouajjani et al.}

\institute{Universit\'e Paris Cit\'e, IRIF, CNRS, Paris, France \\
\email{\{abou,boutglay,haberm\}@irif.fr} \and
Mohammed VI Polytechnic University, Ben Guerir, Morocco}

\iflong
\SetWatermarkAngle{0}
\SetWatermarkText{\raisebox{12.5cm}{
\hspace{-0.13cm}
\href{https://doi.org/10.5281/zenodo.15221107}{\includegraphics{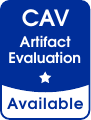}}
\hspace{8.65cm}
\includegraphics{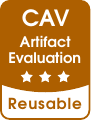}
}}
\else
\SetWatermarkAngle{0}
\SetWatermarkText{\raisebox{11.5cm}{
\hspace{-0.13cm}
\href{https://doi.org/10.5281/zenodo.15221107}{\includegraphics{1-available}}
\hspace{8.65cm}
\includegraphics{3-reusable}
}}
\fi

\maketitle
\begin{abstract}
We address the problem of verifying automatically procedural programs manipulating parametric-size arrays of integers, encoded as a constrained Horn clauses solving problem. We propose a new algorithmic method for synthesizing loop invariants and procedure pre/post-conditions represented as universally quantified first-order formulas constraining the array elements and program variables. We adopt a data-driven approach that extends the decision tree Horn-ICE framework to handle arrays. We provide a powerful learning technique based on reducing a complex classification problem of \emph{vectors of integer arrays} to a simpler classification problem of \emph{vectors of integers}. The obtained classifier is generalized to get universally quantified invariants and procedure pre/post-conditions. We have implemented our method and shown its efficiency and competitiveness w.r.t. state-of-the-art tools on a significant benchmark.

\end{abstract}

% 1. Introduction
\section{Introduction}

Automatic verification of procedural programs manipulating arrays is a challenging problem for which methods able to handle large classes of programs in practice are needed. Verifying that a program satisfies its specification given by a pre-condition and a post-condition amounts to synthesizing accurate loop invariants and procedure pre/post-conditions allowing to establish that every computation starting from a state satisfying the pre-condition cannot reach a state violating the post-condition.
The automatic synthesis of invariants and pre/post-conditions has received a lot of interest from the community. It has been addressed using various approaches leading to the development of multiple verification methods and tools~\cite{quantified-predicate-abstraction,quantified-abstract-domain,angluin-based-array-invs,lazy-abstraction-for-arrays,cell-morphing,quic3,freqhorn-quantifiers,data-abstraction,diffy,prophic3}. 

In this work, we address the problem of verifying procedural programs, with loops and potentially recursive procedures, manipulating integer arrays with {\em parametric} sizes, i.e., the sizes of arrays are considered as parameters explicitly mentioned in the program and its specification. We consider specifications written in first-order logic of arrays with linear constraints. Our contribution is to provide a new {\em data-driven} method for solving this verification problem. 
%based on a learning schema~\cite{ice}. 
Our method generates automatically invariants and procedure pre/post-conditions of
%complex
programs expressed as universally quantified formulas over arrays with integer data.

We use a learning approach for inductive invariant synthesis that consists in taking the set of all program states as the universe and considering two classes: the states reachable from the pre-condition are classified as {\em positive}, and those that can reach states violating the post-condition are classified as {\em negative}. A learner proposes a candidate invariant $I$ to a teacher who checks that (1) the pre-condition is included in $I$, (2) $I$ is included in the post-condition, and (3) $I$ is inductive (i.e., stable under execution of program actions). 
If condition (1), resp. (2), is not satisfied, the teacher provides counterexamples to the learner that are positive, resp. negative. If (3) is not satisfied, the teacher cannot provide positive/negative examples, but communicates to the learner {\em implications} of the form $s \rightarrow s'$ meaning that if state $s$ is included in the invariant, then state $s'$ should be in it too. The use of such conditional classification data in the context of invariant learning (to exploit local reachability information) has been introduced in the ICE framework~\cite{ice} and its instance ICE-DT~\cite{ice-dt} where invariants are generated using decision-tree learning techniques.
We follow this approach. Actually, since we consider programs with procedure calls and recursion, our method is based on the Horn-ICE-DT learning schema that generalizes ICE-DT to constrained Horn clauses~\cite{horn-ice}. Horn-ICE has been applied previously to programs manipulating numerical variables, but never to procedural programs with integer arrays.

To adopt the Horn-ICE-DT schema, one needs to define a learner and a teacher. For the teacher, we use simply the Z3~\cite{z3} solver which can handle array logics~\cite{array-theory,erp-in-z3,arrays-in-z3}. Our contribution is a new decision-tree based learning method that can generate universally quantified first-order formulas on arrays with integers. 

To define the learner, a crucial point is to define the space of attributes (predicates) that could be used for building the decision trees. This space depends on the type of the program states and the targeted class of invariants. For the programs we consider, states are valuations of program variables, of array bounds, and of the arrays (i.e., values stored in the arrays). Our goal is to learn invariants represented as formulas relating program variables with parametric array bounds, universally quantified index variables, and array elements at the positions given by the index variables. 
Then, one issue to address is, given a consistent sample of program states (i.e. no negative configuration is reachable from a positive one), to determine the number of quantified index variables that are needed for defining a classifier that separates correctly the sample. Once this number is fixed the question is what is the relevant relation that exists between program variables, index variables, and array elements. To tackle these issues, we adopt an approach that iteratively considers increasing numbers of quantifiers, and for each fixed number, reduces the learning problem from the original sample of program states (that includes array valuations) to another learning problem on a sample where elements are vectors of integers. This allows to use integer predicates for building decision trees
which are then converted to universally quantified formulas
corresponding to a classifier for the original problem. 
% and then producing a classifier for the original problem by converting these decision trees to universally quantified formulas.
This reduction is nontrivial and requires to define a tight relation between the two learning problems. Roughly, given a consistent sample $\mathcal{S}$ of array-based data points, our method is able to determine the number $n$ which is sufficient for its classification, and to generate a classifier for it from the classifier of another sample $\mathcal{S}'_n$ on integer-based data points.

The learning method we have defined allows to discover complex invariants and procedures' pre/post-conditions of programs with integer arrays that cannot be generated by existing tools for array program verification.
%Figure~\ref{fig:horn-ice-arch} summarizes our method.
We implemented our method and conducted experiments with a large and diverse benchmark of iterative and recursive programs, including array programs from SV-COMP.
%They implement standard algorithms over arrays including different sorting algorithms.
Experimental results show that, within a 300s timeout, our tool verifies more instances than existing tools \textsc{Spacer}~\cite{spacer,quic3}, \textsc{Ultimate Automizer}~\cite{uautomizer} \textsc{FreqHorn}~\cite{freqhorn,freqhorn-quantifiers}, \textsc{Vajra}~\cite{vajra}, \textsc{Diffy}~\cite{diffy}, \textsc{Rapid}~\cite{rapid}, \textsc{Prophic3}~\cite{prophic3} and \textsc{MonoCera}\cite{monocera} with competitive efficiency.
Moreover, our tool can verify recursive procedure programs, which are beyond the capabilities of \textsc{FreqHorn}, \textsc{Vajra}, \textsc{Diffy}, \textsc{Rapid} and \textsc{Prophic3}.

\ssubsubsection{Related work}
Numerous techniques have been developed to infer quantified invariants, which are essential for the verification of array-manipulating programs with parametric sizes.
Predicate abstraction~\cite{predicate-abstraction-1,predicate-abstraction-2} was extended through the use of skolem variables to support quantified reasoning~\cite{quantified-predicate-abstraction}.
Safari~\cite{safari} implements lazy abstraction with interpolants~\cite{lazy-abstraction} tailored for arrays~\cite{lazy-abstraction-for-arrays}.
It was augmented with acceleration techniques in Booster~\cite{booster}.
Abstract domains~\cite{cousot-arrays,quantified-abstract-domain} were introduced for building abstract interpreters~\cite{abstract-interpretation} of programs with arrays.
\cite{quantified-abstract-domain}~leverages existing quantifier-free domains for handling universally quantified properties.
Full-program induction of~\cite{vajra} allows proving quantified properties within a restricted class of array programs expanded in Diffy~\cite{diffy} using difference invariants.
Many of these approaches are limited to non-recursive programs
corresponding to linear CHC.
This is not a limitation for our method or for \textsc{Quic3}~\cite{quic3} and its integration within \textsc{Spacer}~\cite{spacer} that extends IC3/PDR~\cite{ic3,pdr} to non-linear CHC.
A prior work to \textsc{Quic3} is UPDR~\cite{updr} extending IC3 to infer universally quantified invariants for programs modeled using EPR.

There are multiple learning-based methods for invariant synthesis
\cite{ice,loopinvgen,overfitting-in-synthesis,freqhorn,trace-logic}.
Several have been extended to quantified invariants
\cite{freqhorn-quantifiers,rapid}. 
Some of these rely on user-provided templates for invariants~\cite{inv-synth-for-theo,angluin-based-array-invs}.
FreqHorn~\cite{freqhorn}, a notable CHC solver for quantified invariants, combines data-guided syntax synthesis with range analysis, but faces limitations in handling complex iterating patterns or recursive calls (non-linearity of CHC).
This is also the case for~\cite{maxpranq} tailored to the inference of weakest preconditions of linear array programs.
We overcome these issues by using a fully data-driven approach and decision trees for learning successfully used in some ICE instantiations~\cite{ice-dt,boutglay22}.
While ICE has been instantiated previously for learning quantified data automata~\cite{quantified-data-automata} as invariants for linear data structures like singly-linked lists, our method instantiates the more general Horn-ICE framework~\cite{horn-ice}, allowing verification of programs with arbitrary control flow structures and procedure calls.
HoIce~\cite{hoice} is another extension of ICE to solve CHC problems.
While in our work, we adopt the terminology of Horn-ICE and instantiate it, our method is also applicable within the HoIce framework.
Prior instantiations of Horn-ICE were restricted to verifying numerical programs, while HoIce offers only basic support for arrays and lacks the ability to infer quantified invariants.

In \cite{fo-quantified-sep} an algorithm that learns a formula with quantifier-alternation separating positive and negative models is given. But, similar to UPDR, it requires programs modeled using EPR.
Reducing a classification problem with array values to this formalism might be possible but is not straightforward.
\textsc{Rapid}~\cite{rapid} can also learn invariants with quantifier alternation, but is restricted to non-recursive programs with simple control-flow.
%structures.

Alternatively, array programs can be solved without inferring quantified invariants by reducing the safety problem to one with arrays abstracted to a fixed number of variables~\cite{cell-morphing,array-loop-shrinking,viap,data-abstraction}, however the resulting program may be challenging to verify~\cite{julien-braine-thesis}.
\textsc{Prophic3}~\cite{prophic3} mitigates this challenge by combining this abstraction technique with counterexample-guided abstraction refinement within the IC3/PDR framework.
Moreover, it has the capability to reconstruct the quantified invariants for the original system.
Similarly, \textsc{Lambda}~\cite{lambda} is designed for the verification of parametric systems and leverages \textsc{ic3ia}~\cite{ic3ia} as a quantifier-free model checker.
\textsc{MonoCera}~\cite{monocera} simplifies the verification problem by instrumenting the program without eliminating the arrays through abstraction.
In contrast, our method preserves the original verification problem and applies reduction solely to at the level of the learner, transforming the inference of quantified invariants into a scalar classification task.

% Overview
\section{Overview}
\label{sec:overview}
We demonstrate our method by applying it to the program
in Fig.~\ref{fig:bubble-sort} implementing the bubble sort algorithm over an integer array~$a$ (line~\ref{bsx:line:declare-array}) with parametric size~$N$ (line~\ref{bsx:line:declare-parameter})\footnote{We adhere to standard C semantics, which stipulate that stack-allocated variables, if uninitialized, may assume arbitrary values.}.
The precondition of the program is given by the~\texttt{assume} statement, and its postcondition is given by the ~\texttt{assert} statement in line~\ref{bsx:line:assert} (verifying whether the array~$a$ at this point is a permutation of the initial array falls outside the scope of this paper).

\begin{figure}[ht]
  \centering
  \scriptsize
  \begin{minipage}{0.41\textwidth}
    \begin{lstlisting}[numbers=left,breaklines=true,escapechar=$]
void main() {
  unsigned int N;$\label{bsx:line:declare-parameter}$
  assume($\mathcolorbox{orange!50}{\text{N > 0}}$);$\label{bsx:line:assume}$
  int a[$\mathcolorbox{brown!50}{\text{N}}$];$\label{bsx:line:declare-array}$
  bool $\mathcolorbox{brown!25}{\text{s = true}}$;
  while($\mathcolorbox{red!75}{\text{s}}$) {
    $\mathcolorbox{red!50}{\text{s = false}}$;
    unsigned int $\mathcolorbox{red!25}{\text{i = 1}}$;
    while($\mathcolorbox{purple!75}{\text{i < N}}$) {
     \end{lstlisting}
  \end{minipage}
  \begin{minipage}{0.41\textwidth}
    \begin{lstlisting}[firstnumber=10,numbers=left,breaklines=true,escapechar=$]
      if($\mathcolorbox{cyan!75}{\text{a[i - 1] > a[i]}}$) {
        $\mathcolorbox{cyan!50}{\text{int tmp = a[i]}}$;
        $\mathcolorbox{cyan!50}{\text{a[i] = a[i - 1]}}$;
        $\mathcolorbox{cyan!50}{\text{a[i - 1] = tmp}}$;
        $\mathcolorbox{cyan!25}{\text{s = true}}$;      }
      $\mathcolorbox{purple!25}{\text{i++}}$;} }
  assert($\mathcolorbox{orange!75}{\forall k_1, k_2.\; 0 \leq k_1 \leq k_2 < N}$$\label{bsx:line:assert}$
              $\mathcolorbox{orange!75}{\implies a[k_1] \leq a[k_2]}$);}
    \end{lstlisting}
  \end{minipage}
  \caption{Bubble sort over a parametric size array of integers.}
  \label{fig:bubble-sort}
\end{figure}

We start by reducing the safety verification of the program to the satisfiability of a system of
constrained Horn clauses (CHC). This is achieved using the methodology described e.g. in~\cite{chc-for-verification-1,chc-for-verification-2}.
For our example, the corresponding system is given below, where all free variables in each clause are implicitly universally quantified.
{\small
  \begin{align}
\mathcolorbox{orange!50}{N > 0} \land \mathcolorbox{brown!50}{\BoundedArrayLength{a} = N} \land \mathcolorbox{brown!25}{s} & \implies \Predicate{I_0}(N, a, s) \\
\Predicate{I_0}(N, a, s) \land \mathcolorbox{red!75}{s} \land \mathcolorbox{red!50}{\neg s'} \land \mathcolorbox{red!25}{i = 1} & \implies \Predicate{I_1}(N, a, s', i) \\
\Predicate{I_1}(N, a, s, i) \land \mathcolorbox{purple!75}{i < N} \land \neg(\mathcolorbox{cyan!75}{\BoundedArrayRead{a}{i - 1} > \BoundedArrayRead{a}{i}}) \land \mathcolorbox{purple!25}{i' = i + 1} & \implies \Predicate{I_1}(N, a, s, i') \\
\Predicate{I_1}(N, a, s, i) \land \mathcolorbox{purple!75}{i < N} \land \mathcolorbox{cyan!75}{\BoundedArrayRead{a}{i - 1} > \BoundedArrayRead{a}{i}} \land \mathcolorbox{cyan!25}{s'} \land \mathcolorbox{purple!25}{i' = i + 1} & \notag \\ \land \mathcolorbox{cyan!50}{a' = \BoundedArrayWrite{a}{i}{\BoundedArrayRead{a}{i - 1}} \land a'' = \BoundedArrayWrite{a'}{i - 1}{\BoundedArrayRead{a}{i}}} & \implies \Predicate{I_1}(N, a'', s', i') \\
\Predicate{I_1}(N, a, s, i) \land \neg(\mathcolorbox{purple!75}{i < N}) & \implies \Predicate{I_0}(N, a, s) \\
\Predicate{I_0}(N, a, s) \land \neg \mathcolorbox{red!75}{s} & \notag \\ \land \neg (\mathcolorbox{orange!75}{\forall k_1, k_2.\; 0 \leq k_1 \leq k_2 < N \implies a[k_1] \leq a[k_2]}) & \implies \bot
  \end{align}
}
The system above defines constraints on the set of uninterpreted predicates~$\mathcal{P} = \{\Predicate{I_0}, \Predicate{I_1}\}$, where $\Predicate{I_0}$~and~$\Predicate{I_1}$ represent the invariants of the outer loop \lstinline[columns=fixed]{while(s)} and the nested loop \lstinline[columns=fixed]{while(i < N)}, respectively.
In these constraints, $\BoundedArrayRead{a}{i}$ represents the value of the array~$a$ at index~$i$ and $\BoundedArrayWrite{a}{i}{v}$ represents an array with the same length and elements as~$a$, except at index~$i$ where it has the value~$v$.
The program in Fig.~\ref{fig:bubble-sort} is safe if and only if this system is satisfiable, i.e., there are 
 interpretations for~$\Predicate{I_0}$ and~$\Predicate{I_1}$ satisfying all the clauses.
As we will see below, expressing such interpretations requires using universally quantified first-order formulas.
%, and it is the case when solving most of the systems corresponding to programs with arrays. 

Our method to solve CHCs like the one above is based on Horn-ICE~\cite{horn-ice} learning approach which follows the standard learning loop where a learner and a teacher interact iteratively, the learner using a sample (set of examples) to infer a candidate solution and a teacher either approving it when a solution is found, or otherwise providing counterexamples that can be used in the next learning iteration. Horn-ICE is an extension of this principle that is adapted to learning inductive invariants by using, in addition to positive and negative examples, implications that provide conditional information such as: if some states are in the invariant, then necessarily some other state must also be in the invariant.
Let us describe briefly this schema. Consider a CHC system built from a program as in the example above. In each iteration, the learner generates for each uninterpreted predicate in the system an interpretation using a \emph{data point sample} $\mathcal{S} = (X, C)$ where $X$ is a set of data points and $C$ a set of Horn implications over $X$.
A \emph{data point} $x \in X$ corresponds to a configuration of the program at some location.
%or input/output configurations for a postcondition.
To each data point is assigned an uninterpreted predicate $\Predicate{P} \in \mathcal{P}$ (denoted by $\PredicateOfDatapoint{x}$) and a vector of constants, one for each parameter variable of $\Predicate{P}$.
A \emph{Horn implication} over $X$ is a hyper-edge (generalizing ICE's implication) of one of the following forms:
(a) $\top \rightarrow x$, where $x \in X$, meaning $x$ should satisfy
  the predicate $\PredicateOfDatapoint{x}$;
(b) $x_1 \land  \dots \land x_n \rightarrow x$, where $x_1,\dots,x_n,x \in X$, i.e. if $x_1,\dots,x_n$, respectively, satisfy $\PredicateOfDatapoint{x_1},\dots,\PredicateOfDatapoint{x_n}$ then $x$ should satisfy $\PredicateOfDatapoint{x}$, or if $x$ doesn't satisfy $\PredicateOfDatapoint{x}$ at least one of the $x_1,\dots,x_n$ should not satisfy its predicate;
(c) $x_1 \land  \dots \land x_n \rightarrow \bot$, where $x_1,\dots,x_n \in X$,  i.e. at least one of the $x_1,\dots,x_n$ should not satisfy its predicate.
  A data point sample $\mathcal{S} = (X, C)$ is called {\em consistent} if it
  admits a {\em consistent labeling} which
  labels each element $x$ of $X$ with either $\top$ or $\bot$
  while satisfying all the constraints in $C$.

The teacher checks if the generated interpretations by the learner satisfy the CHC system and provides feedback. 
% validity of the candidate solution by considering the satisfiability of each clause by inserting for each predicate in $\mathcal{P}$ the interpretation given by the candidate solution.
If a clause $\forall \vec{v}.\; \phi(\vec{v}) \implies \Predicate{P_j}(\vec{v})$ is violated, then a counterexample is computed which is a data point $x$ associated with the predicate $\Predicate{P_j} \in \mathcal{P}$ together with a Horn implication $\top \rightarrow x$.
If a clause $\forall \vec{v}_1,\dots,\vec{v}_n.\; \Predicate{P_1}(\vec{v}_1) \land \dots \land \Predicate{P_n}(\vec{v}_n) \land \phi(\vec{v}_1,\dots,\vec{v}_n) \implies \bot$ is violated, the counterexample is data points $x_1,\dots,x_n$ with Horn implication $x_1 \land \dots \land x_n \rightarrow \bot$.
If a clause $\forall \vec{v}_1,\dots,\vec{v}_n,\vec{v}.\; \Predicate{P_1}(\vec{v}_1) \land \dots \land \Predicate{P_n}(\vec{v}_n) \land \phi(\vec{v}_1,\dots,\vec{v}_n,\vec{v}) \implies \Predicate{P_j}(\vec{v})$ is violated, the counterexample is data points $x_1,\dots,x_n, x$ with Horn implication $x_1 \land \dots \land x_n \rightarrow x$.

To make this schema work, one has to define a learner and a teacher, depending on the considered classes of programs and properties. In this paper, we apply this schema to handle programs with (parametric-size) arrays and properties expressed in first-order logic of arrays, which has not been done so far. 
For the teacher, we rely on using the Z3~\cite{z3} SMT solver which can handle different decidable fragments of array logics~\cite{array-theory,erp-in-z3,arrays-in-z3}\footnote{In the literature arrays are typically handled using uninterpreted functions. We can easily encode parametric-size arrays like that.}.
Z3 attempts in addition to solve queries beyond the known decidable fragments using various heuristics.
Then, our main contribution consists in providing a new learning technique able to synthesize invariants/procedure summaries as universally quantified formulas over arrays. This requires addressing a number of nontrivial problems. Let us first see how the learner and the teacher interact, and what is the type of information they exchange, in the case of the bubble-sort example (Fig.~\ref{fig:bubble-sort}).

%For example, during the verification of Fig.~\ref{fig:bubble-sort} using our method, 
In the first iteration, starting with an empty sample (with no counterexamples), the learner proposes $\Predicate{I_0}$ and $\Predicate{I_1}$ as \texttt{true}.
The teacher finds this violates clause~(6) and provides the counterexample $\langle \Predicate{I_0}, N \mapsto 2, a \mapsto [1, 0], s \mapsto \bot \rangle \rightarrow \bot$, indicating that this data point must not be included in $\Predicate{I_0}$'s invariant (exiting the outer loop while $a$ is not sorted).
In the second iteration, the learner proposes for $\Predicate{I_0}$ $\forall k_1, k_2.\; 0 \leq k_1 \leq k_2 < \BoundedArrayLength{a} \implies a[k_1] > 0$ and keeps \texttt{true} for $\Predicate{I_1}$.
The teacher identifies a violation of clause~(1) and provides the counterexample $\top \rightarrow \langle \Predicate{I_0}, N \mapsto 1, a \mapsto [0], s \mapsto \top \rangle$, indicating this configuration must be included in $\Predicate{I_0}$'s invariant as it is a valid initial state.
After collecting more counterexamples, the learner proposes $\forall k_1, k_2.\; 0 \leq k_1 \leq k_2 < \BoundedArrayLength{a} \implies a[k_1] \leq 0 \lor s$ for $\Predicate{I_0}$ and \texttt{true} for $\Predicate{I_1}$. The teacher then reports a violation of clause~(5) with the implication counterexample $\langle \Predicate{I_1}, N \mapsto 1, a \mapsto [1], i \mapsto 1, s \mapsto \bot \rangle \rightarrow \langle \Predicate{I_0}, N \mapsto 1, a \mapsto [1], s \mapsto \bot \rangle$ indicating that if the first configuration is in $\Predicate{I_1}$, the second should also be in $\Predicate{I_0}$.

\begin{figure}
\centering
\begin{subfigure}[b]{0.49\textwidth}
\scalebox{0.7}{
  \begin{tikzpicture}
    \node[anchor=east] (0) at (-3.5, -1) {$\top$};

    \node[anchor=east] (11) at (0, 0) {$\langle \Predicate{I_0}, 1, [0], \top \rangle^1$};
    
    \node[anchor=east] (21) at (0, -0.5) {$\langle \Predicate{I_0}, 1, [1], \top \rangle^2$};

    \node[anchor=east] (31) at (0, -1.5) {$\langle \Predicate{I_0}, 2, [0, 0], \top \rangle^{\star}$};
    
    \node[anchor=east] (41) at (0, -2) {$\langle \Predicate{I_0}, 2, [1, 0], \top \rangle^{\dagger}$};
    \node[anchor=east] (42) at (4, -2) {$\langle \Predicate{I_1}, 2, [1, 0], 1, \bot\rangle^{\ddagger}$};

    \node[anchor=east] (61) at (4, -1) {$\langle \Predicate{I_1}, 1, [1], 1, \bot\rangle^4$};
    \node[anchor=east] (62) at (0, -1) {$\langle \Predicate{I_0}, 1, [1], \bot \rangle^3$};

    \node[anchor=east] (71) at (4, -2.5) {$\langle \Predicate{I_1}, 2, [0, 0], 2, \bot \rangle^\circ$};
    \node[anchor=east] (72) at (0, -2.5) {$\langle \Predicate{I_0}, 2, [0, 0], \bot \rangle^\diamond$};

    \node[anchor=east] (51) at (0, -3) {$\langle \Predicate{I_0}, 2, [1, 0], \bot \rangle^\triangleleft$};
    \node[anchor=east] (52) at (4, -3) {$\langle \Predicate{I_1}, 2, [1, 0], 2, \bot\rangle^\triangleright$};

    \node[anchor=east] (1) at (-3.5, -3) {$\bot$};

    \draw[->, tips=on proper draw] (0.east) edge (11.west);
    \draw[->, tips=on proper draw] (0.east) edge (21.west);
    \draw[->, tips=on proper draw] (0.east) edge (31.west);

    \draw[->, tips=on proper draw] (0.east) edge (41.west);
    \draw[->, tips=on proper draw] (41.east) edge (42.west);

    \draw[->, tips=on proper draw] (51.west) edge (1.east);
    \draw[->, tips=on proper draw] (52.west) edge (51.east);

    \draw[->, tips=on proper draw] (61.west) edge (62.east);
    \draw[->, tips=on proper draw] (71.west) edge (72.east);
  \end{tikzpicture}
}
   \caption{}
   \label{fig:data-sample} 
\end{subfigure}
\begin{subfigure}[b]{0.49\textwidth}
\scalebox{0.7}{
  \begin{tikzpicture}
    \node[anchor=east] (0) at (-4, -1.5) {$\top$};

    \node[anchor=east] (11) at (0, 0) {$\langle \Predicate{I_0}, 1, \top, 0, 0, 0, 0, 1 \rangle^1$};
    
    \node[anchor=east] (21) at (0, -0.5) {$\langle \Predicate{I_0}, 1, \top, 0, 0, 1, 1, 1 \rangle^2$};

    \node[anchor=east] (311) at (0, -1.5) {$\langle \Predicate{I_0}, 2, \top, 0, 0, 0, 0, 2 \rangle^{\star}$};
    \node[anchor=east] (312) at (0, -2) {$\langle \Predicate{I_0}, 2, \top, 0, 1, 0, 0, 2 \rangle^{\star}$};
    \node[anchor=east] (313) at (0, -2.5) {$\langle \Predicate{I_0}, 2, \top, 1, 1, 0, 0, 2 \rangle^{\star\dagger}$};
    
    \node[anchor=east] (411) at (0, -3) {$\langle \Predicate{I_0}, 2, \top, 0, 0, 1, 1, 2 \rangle^{\dagger}$};
    \node[anchor=east] (412) at (0, -3.5) {$\langle \Predicate{I_0}, 2, \top, 0, 1, 1, 0, 2 \rangle^{\dagger}$};
    \node[anchor=east] (421) at (4.5, -2.5) {$\langle \Predicate{I_1}, 2, 1, \bot, 0, 0, 1, 1, 2 \rangle^{\ddagger}$};
    \node[anchor=east] (422) at (4.5, -3) {$\langle \Predicate{I_1}, 2, 1, \bot, 0, 1, 1, 0, 2 \rangle^{\ddagger}$};
    \node[anchor=east] (423) at (4.5, -3.5) {$\langle \Predicate{I_1}, 2, 1, \bot, 1, 1, 0, 0, 2 \rangle^{\ddagger}$};

    \node[anchor=east] (61) at (4.5, -1) {$\langle \Predicate{I_1}, 1, 1, \bot, 0, 0, 1, 1, 1 \rangle^4$};
    \node[anchor=east] (62) at (0, -1) {$\langle \Predicate{I_0}, 1, \bot, 0, 0, 1, 1, 1 \rangle^3$};

    \node[anchor=east] (711) at (4.5, -4) {$\langle \Predicate{I_1}, 2, 2, \bot, 0, 0, 0, 0, 2 \rangle^\circ$};
    \node[anchor=east] (712) at (4.5, -4.5) {$\langle \Predicate{I_1}, 2, 2, \bot, 0, 1, 0, 0, 2 \rangle^\circ$};
    \node[anchor=east] (713) at (4.5, -5) {$\langle \Predicate{I_1}, 2, 2, \bot, 1, 1, 0, 0, 2 \rangle^{\circ\triangleright}$};
    \node[anchor=east] (721) at (0, -4) {$\langle \Predicate{I_0}, 2, \bot, 0, 0, 0, 0, 2 \rangle^\diamond$};
    \node[anchor=east] (722) at (0, -4.5) {$\langle \Predicate{I_0}, 2, \bot, 0, 1, 0, 0, 2 \rangle^\diamond$};
    \node[anchor=east] (723) at (0, -5) {$\langle \Predicate{I_0}, 2, \bot, 1, 1, 0, 0, 2 \rangle^{\diamond\triangleleft}$};

    \node[anchor=east] (511) at (0, -5.5) {$\langle \Predicate{I_0}, 2, \bot, 0, 0, 1, 1, 2 \rangle^\triangleleft$};
    \node[anchor=east] (512) at (0, -6) {$\langle \Predicate{I_0}, 2, \bot, 0, 1, 1, 0, 2 \rangle^\triangleleft$};
    \node[anchor=east] (521) at (4.5, -5.5) {$\langle \Predicate{I_1}, 2, 2, \bot, 0, 0, 1, 1, 2 \rangle^\triangleright$};
    \node[anchor=east] (522) at (4.5, -6) {$\langle \Predicate{I_1}, 2, 2, \bot, 0, 1, 1, 0, 2 \rangle^\triangleright$};

    \node[anchor=east] (1) at (-4, -5.5) {$\bot$};

    \draw[->, tips=on proper draw] (0.east) edge (11.west);
    \draw[->, tips=on proper draw] (0.east) edge (21.west);
    
    \draw[->, tips=on proper draw] (0.east) edge (311.west);
    \draw[->, tips=on proper draw] (0.east) edge (312.west);
    \draw[->, tips=on proper draw] (0.east) edge (313.west);

    \draw[->, tips=on proper draw] (0.east) edge (411.west);
    \draw[->, tips=on proper draw] (0.east) edge (412.west);
    \draw[->, tips=on proper draw] (313.east) -- ++(0.75, 0) coordinate (C41) -- (421.west);
    \draw (411.east) -- (C41);
    \draw (412.east) -- (C41);
    \draw[fill=white] (C41) circle (1.1pt);
    \draw[->, tips=on proper draw] (411.east) -- ++(0.75, 0) coordinate (C42) -- (422.west);
    \draw (313.east) -- (C42);
    \draw (412.east) -- (C42);
    \draw[fill=white] (C42) circle (1.1pt);
    \draw[->, tips=on proper draw] (412.east) -- ++(0.75, 0) coordinate (C43) -- (423.west);
    \draw (411.east) -- (C43);
    \draw (313.east) -- (C43);
    \draw[fill=white] (C43) circle (1.1pt);

    \draw[->, tips=on proper draw] (711.west) -- ++(-0.75, 0) coordinate (C71) -- (721.east);
    \draw (712.west) -- (C71);
    \draw (713.west) -- (C71);
    \draw[fill=white] (C71) circle (1.1pt);
    \draw[->, tips=on proper draw] (712.west) -- ++(-0.75, 0) coordinate (C72) -- (722.east);
    \draw (711.west) -- (C72);
    \draw (713.west) -- (C72);
    \draw[fill=white] (C72) circle (1.1pt);
    \draw[->, tips=on proper draw] (713.west) -- ++(-0.75, 0.25) coordinate (C73) -- (723.east);
    \draw (711.west) -- (C73);
    \draw (712.west) -- (C73);
    \draw[fill=white] (C73) circle (1.1pt);

    \draw[->, tips=on proper draw] (713.west) -- ++(-0.75, -0.25) coordinate (C51) -- (723.east);
    \draw (521.west) -- (C51);
    \draw (522.west) -- (C51);
    \draw[fill=white] (C51) circle (1.1pt);
    \draw[->, tips=on proper draw] (521.west) -- ++(-0.75, 0) coordinate (C52) -- (511.east);
    \draw (713.west) -- (C52);
    \draw (522.west) -- (C52);
    \draw[fill=white] (C52) circle (1.1pt);
    \draw[->, tips=on proper draw] (522.west) -- ++(-0.75, 0) coordinate (C53) -- (512.east);
    \draw (713.west) -- (C53);
    \draw (521.west) -- (C53);
    \draw[fill=white] (C53) circle (1.1pt);

    \draw[->, tips=on proper draw] (511.west) -- ++(-0.25, 0) coordinate (CC3) -- (1.east);
    \draw (723.west) -- (CC3);
    \draw (512.west) -- (CC3);
    \draw[fill=white] (CC3) circle (1.1pt);

    \draw[->, tips=on proper draw] (61.west) edge (62.east);
  \end{tikzpicture}
}
   \caption{}
   \label{fig:diagram-sample}
\end{subfigure}
\caption{(a) A data point sample during verification of bubble sort and (b) its diagram sample using 2 quantifier variables.
Diagrams of (b) are derived from data points with the same exponent in (a).}
\label{fig:ns}
\end{figure}
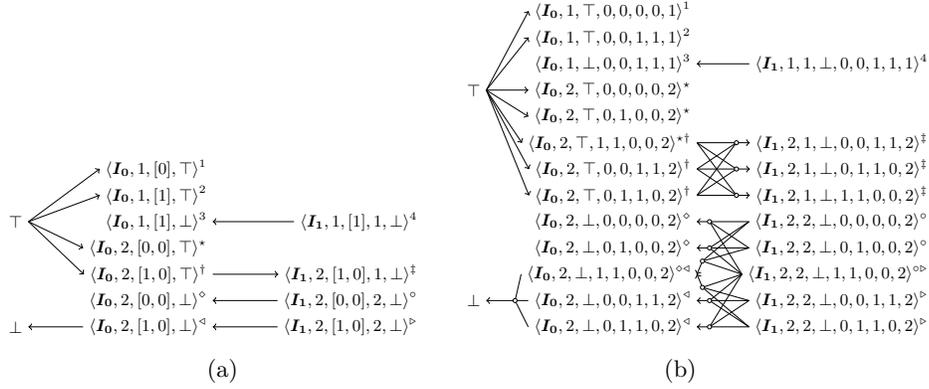

Now, the question is how the learner synthesizes a candidate solution from a given consistent data point sample $\mathcal{S} = (X, C)$. The general principle is to build a formula that is a {\em classifier} of  $\mathcal{S}$, i.e., that separates the elements of $X$ into positive and negative ones while respecting the constraints in $C$ (collected from counterexamples to inductiveness during the learning process). The challenge is to generate universally quantified formulas over (parametric-size) arrays from a finite set of data, and for that there are two important questions: (1) how many quantifiers are needed to express the solution? and (2) what is the mechanism to use to generate the constraints on indexed elements of arrays and program variables? 

Let us keep the first question for later, and assume for the moment that the number of quantifiers is given. To address the second question, we adopt a learning mechanism based on decision-trees following the Horn-ICE-DT schema \cite{horn-ice}, which is a natural approach for generating formulas. Then, the crucial questions are what are the predicates to use as attributes to check at the nodes of these decision trees? and how to use these predicates to build a universally quantified formula? 
%Answering to these questions constitutes actually our main conceptual contribution in this work. 
To make the space of the possible predicates easier to explore, we reduce our classification problem on array-based data points samples to another classification problem stated on integer-based data points samples that can be solved using integer constraints. 
%The reduction allows to leverage a classifier of the integer-based sample expressed as a formula on a number of elements at some parametric positions in the arrays, to define a classifier of the original array-based sample using universal quantification over the positions of these elements. 

In more details, we introduce a technique called {\em diagramization} summarized as follows:
Given a classification problem of a consistent sample $\mathcal{S}$ of program states including array valuations (assume that we have one array $a$ to simplify the explanation), a fixed number $n$ and a set $V$ of size $n$ of fresh index variables to be universally quantified in the classifier (called {\em quantifier variables}), we consider another classification problem on a sample $\mathcal{S}'_n$ of so-called {\em diagrams}. They
associate values to program variables, to variables $k$ in $V$, and to terms $a[k]$ (representing the element of $a$ at position $k$). Elements of $\mathcal{S}'_n$ are vectors of integers obtained from elements of $\mathcal{S}$ by taking all possible projections of arrays on a fixed number of positions.  
The sample $\mathcal{S}'_n$ is obtained by considering for each array valuation in a state, all possible mappings from $V$ to its array elements. Moreover, we transfer the classification information of $\mathcal{S}$ to $\mathcal{S}'_n$.
%For that, we must impose the following facts: For every element $s \in \mathcal{S}$, if $s$ is classified positively, then all the elements in $\mathcal{S}'_n$ obtained from $s$ must be positive.  However, if $s$ is classified negatively, then at least one of the elements obtained from $s$ should be negative. 
%Notice that to express the latter fact, we need to use disjunctive classification constraints (not only simple classification facts of elements as positive or negative). 
%This step in our method is called {\em diagramization}. 

At this point, a question is whether $\mathcal{S}'_n$ is consistent (knowing that $\mathcal{S}$ is consistent). 
Let us assume it is for the moment and come back to this issue later.
Then, the learner proceeds by constructing a decision tree for $\mathcal{S}'_n$ using predicates on integers as attributes. In our implementation, we use predicates appearing in the program and the specification, as well as predicates generated by progressive enumeration from simple patterns in domains such as interval or octagonal constraints~\cite{octagon}. In fact, it is possible to determine if a given set of attributes allows to build a classifier of a given sample, and if it is does not, to generate additional attributes from the considered patterns (a sufficient set of attributes is guaranteed to be found for a consistent sample). 
%Eventually a sufficient set of attributes is found since the considered patterns allow in the worst case to characterize individual data points. 
Then, we prove that when a classifier is found for $\mathcal{S}'_n$ (expressed as a formula relating program variables, index variables $k$ and corresponding terms $a[k]$), its conversion by universally quantifying over all the $V$ variables is indeed a classifier for $\mathcal{S}$ (see Theorem~\ref{theorem1}). 

%Conversely, if $\mathcal{S}$ is a consistent sample admitting a classifier in our logic, then, necessarily there is a number $n$ of index variables for which there exists a classifier for the sample $\mathcal{S}'_n$.
%
%Otherwise, the failure is due to either the insufficiency of attributes or to the number $n$ of index variables in $V$. 
%If it is the former, more attributes need to be enumerated until they are sufficient. This is guaranteed to work with integer vectors even when enumerating constraints in inexpressive domains such as intervals (i.e., lower and upper bound constraints). If the issue is related to the number $n$, then it is incremented ($V$ is extended by one extra fresh variable), and the procedure is repeated.

%The learner reduces the classification of a data point sample $\mathcal{S}$ (of program's configurations with arrays) to an array-free classification problem $\mathcal{S}'$.
%This new classification problem can be effectively solved using a decision-tree algorithm and its solutions can then be transformed into quantified solutions for the original sample $\mathcal{S}$.
%This reduction is parameterized by the number of quantified variables to use per array.
%Determining this number is a crucial question.

%We now illustrate how the learner in our method synthesizes quantified solutions.
Let us illustrate this process on our bubble-sort example. At the 10th iteration of the verification of Fig.~\ref{fig:bubble-sort}, the learner has accumulated counterexamples shown in the data point sample in Fig.~\ref{fig:data-sample}.
For simplicity, variable names are omitted; e.g., $\langle \Predicate{I_0}, N \mapsto 2, a \mapsto [1, 0], s \mapsto \bot \rangle$ is shortened to $\langle \Predicate{I_0}, 2, [1, 0], \bot \rangle$.
As explained above, the key idea of our method is that the learner's sample can be reduced to a diagram sample, where examples consist only of scalar and boolean values.
For example, for the data point $x_1 = \langle \Predicate{I_0}, N \mapsto 2, a \mapsto [1, 0], s \mapsto \bot \rangle$, if we abstract the array $a$ using two quantifier variables $k_1$ and $k_2$, we introduce fresh variables $a_{k_1}$ and $a_{k_2}$, representing the values of $a$ at the positions indexed by $k_1$ and $k_2$, respectively.
We also introduce an additional fresh variable $l_a$ to represent the size of the array.
We explain later how the number of quantifier variables (2 in this case) is determined.
In this case, $x_1$ is transformed into the following diagrams:\\
\hspace*{6ex}$d_1 = \langle \Predicate{I_0}, N \mapsto 2, s \mapsto \bot, k_1 \mapsto 0, k_2 \mapsto 0, a_{k_1} \mapsto 1, a_{k_2} \mapsto 1, l_{a} \mapsto 2 \rangle$\\
\hspace*{6ex}$d_2 = \langle \Predicate{I_0}, N \mapsto 2, s \mapsto \bot, k_1 \mapsto 0, k_2 \mapsto 1, a_{k_1} \mapsto 1, a_{k_2} \mapsto 0, l_{a} \mapsto 2 \rangle$\\
\hspace*{6ex}$d_3 = \langle \Predicate{I_0}, N \mapsto 2, s \mapsto \bot, k_1 \mapsto 1, k_2 \mapsto 0, a_{k_1} \mapsto 0, a_{k_2} \mapsto 1, l_{a} \mapsto 2 \rangle$\\
\hspace*{6ex}$d_4 = \langle \Predicate{I_0}, N \mapsto 2, s \mapsto \bot, k_1 \mapsto 1, k_2 \mapsto 1, a_{k_1} \mapsto 0, a_{k_2} \mapsto 0, l_{a} \mapsto 2 \rangle$

%\begin{align*}
%  &d_1 = \langle \Predicate{I_0}, N \mapsto 2, s \mapsto \bot, k_1 \mapsto 0, k_2 \mapsto 0, a_{k_1} \mapsto 1, a_{k_2} \mapsto 1, l_{a} \mapsto 2 \rangle \\
%  &d_2 = \langle \Predicate{I_0}, N \mapsto 2, s \mapsto \bot, k_1 \mapsto 0, k_2 \mapsto 1, a_{k_1} \mapsto 1, a_{k_2} \mapsto 0, l_{a} \mapsto 2 \rangle \\
%  &d_3 = \langle \Predicate{I_0}, N \mapsto 2, s \mapsto \bot, k_1 \mapsto 1, k_2 \mapsto 0, a_{k_1} \mapsto 0, a_{k_2} \mapsto 1, l_{a} \mapsto 2 \rangle \\
%  &d_4 = \langle \Predicate{I_0}, N \mapsto 2, s \mapsto \bot, k_1 \mapsto 1, k_2 \mapsto 1, a_{k_1} \mapsto 0, a_{k_2} \mapsto 0, l_{a} \mapsto 2 \rangle
%\end{align*}

In this newly \textit{diagramized} sample, diagrams of a positive data point must all be classified as positive.
For negative data points, at least one diagram must be classified as negative. This is encoded in the diagram sample with the implication over diagrams $d_1 \land d_2 \land d_3 \land d_4 \rightarrow \bot$.
As the data point $x_1$ in the original sample is negative, at least one of its four diagrams must be classified as negative.
Here, $d_2$ is negative, as it violates the program assertion ($k_1 < k_2$ is true but not $a_{k_1} \leq a_{k_2}$).
For implication counterexamples, if all diagrams of the left-hand side data point are classified as positive, then all diagrams of the right-hand side data point must also be classified as positive.
Similarly, if a diagram of the right-hand side data point is classified as negative, then at least one diagram of the left-hand side data point must also be classified as negative.

Our method reduces the data point sample shown in Fig.~\ref{fig:data-sample} to the diagram sample shown in Fig.~\ref{fig:diagram-sample}.
For brevity, variable names are again omitted, so $\langle \Predicate{I_0}, N \mapsto 2, s \mapsto \bot, k_1 \mapsto 0, k_2 \mapsto 1, a_{k_1} \mapsto 1, a_{k_2} \mapsto 0, l_a \mapsto 2 \rangle$ is shortened to $\langle \Predicate{I_0}, 2, \bot, 0, 1, 1, 0, 2 \rangle$.
Notice that different data points may share some diagrams.

Then, the obtained diagram sample is classified by a decision-tree learning algorithm that produces a quantifier-free formula using attributes generated using the domain of octagonal constraints.
%, i.e., patterns that consists of comparisons of terms of the form $x - y$ or $x+y$ with constants.
%Our method then proceeds to classify this diagram sample using a decision-tree learning algorithm in order to produce a quantifier-free formula over the variable of every invariants except array variables and the introduced free variables using attributes over these variables enumerated from a fixed domain like octagons.
% Another important question is the origin of the attributes used by the decision-tree learning algorithm.
The decision-tree learning procedure over this sample yields $s \lor a_{k_1} \leq a_{k_2}$ for~$\Predicate{I_0}$ and $i \leq k_2 \lor a_{k_1} \leq a_{k_2}$ for~$\Predicate{I_1}$.
When universally quantified, we obtain the solution with $\forall k_1, k_2.\; 0 \leq k_1 \leq k_2 < \BoundedArrayLength{a} \implies a[k_1] \leq a[k_2] \lor s$ for $\Predicate{I_0}$ and $\forall k_1, k_2.\; 0 \leq k_1 \leq k_2 < \BoundedArrayLength{a} \implies a[k_1] \leq a[k_2] \lor i \leq k_2 \lor s$ for~$\Predicate{I_1}$ that the learner proposes to the teacher.
%In Theorem~\ref{theorem1}, we prove that this transformation (substitution and quantification) of the solution of $\mathcal{S}'$ results in a solution for $\mathcal{S}$.

Now, let us go back to the question whether $\mathcal{S}'_n$ is consistent. This question is related to another question we left pending earlier in this section which is how to determine the number of quantified variables $n$. In fact, it can be the case that for a consistent sample $\mathcal{S}$ and an integer $n$, the sample $\mathcal{S}'_n$ is not consistent as illustrated later in Example~\ref{ex:not-enough-quantifiers}. This means that in this case $n$ is not sufficient for defining a formula that classifies $\mathcal{S}$. Therefore, our learner has a loop that increments the number of quantifier variables, starting with one quantifier variable per array, until a consistent diagram sample is obtained. This is guaranteed to succeed, as stated in Theorem~\ref{theorem2}.
% since we can at most choose the number of quantifier variables equal to the size of the largest array in the sample.

Finally, let us mention that when after a number of iterations between the learner and the teacher the obtained sample $\mathcal{S}$ is inconsistent, 
the program does not satisfy its specification.

\iflong
  Fig~\ref{fig:horn-ice-arch} provides a graphical summary of the proposed method and Appendix~\ref{appendix:bubble-sort-iterations} provides the omitted iterations of the example.

  \begin{figure}%[t!h]
  \centering
  \resizebox{\textwidth}{!}{
  \begin{tikzpicture}
    \begin{scope}[font=\sffamily]
      \node[draw, minimum width=25mm, minimum height=4ex, thick] (compiler) at (-4, 0) {Parser/compiler};
      \node[draw, minimum width=25mm, minimum height=4ex, thick] (teacher) at (0, 0) {Teacher};
      \node[draw, minimum width=25mm, minimum height=4ex, thick] (smtsolver) at (0, 2) {SMT Solver};
      \node[draw, minimum width=25mm, minimum height=4ex, thick] (learner) at (5.5, 0) {Learner};
      \node[draw, minimum width=25mm, minimum height=4ex, thick] (dtlearner) at (5.5, 2) {DT Learner};
      \node[draw, minimum width=25mm, minimum height=4ex, thick] (diagrammanager) at (11, 0) {Diagramization};
    \end{scope}
    
    \draw[->, shorten <=1pt] (teacher) edge[bend left] node[text width=20mm, align=right, left] {Validity of \\ SMT formula} (smtsolver);
    \draw[->, shorten <=1pt] (smtsolver) edge[bend left]  node[text width=20mm, align=left, right] {Valid, \\ model} (teacher);
    \draw[->, shorten <=1pt] (learner) edge[bend left] node[text width=30mm, align=right, left] {$\mathcal{S}'_n$} (dtlearner);
    \draw[->, shorten <=1pt] (dtlearner) edge[bend left]  node[text width=20mm, align=left, right] {$\varphi$} (learner);
    \draw[->, shorten <=1pt] (teacher) edge[]  node[text width=30mm, align=center, above] {Data points, \\ Horn implications} (learner);
    \draw[shorten <=1pt] (learner.south west) edge[->, bend left=20]  node[text width=50mm, align=center, below] {Candidate solution \\ $\xi(\varphi) = \forall\varphi'$} (teacher.south east);
    \draw[->, shorten <=1pt] (dtlearner) edge[bend left]  node[text width=20mm, align=left, right] {$\varphi$} (learner);
    \draw[->, shorten <=1pt] (learner) edge[]  node[text width=30mm, align=center, above] {$n$, \\ $\mathcal{S} = (X, C)$} (diagrammanager);
    \draw[shorten <=1pt] (diagrammanager.south west) edge[->, bend left=20]  node[text width=50mm, align=center, below] {$\mathcal{S}'_n = \delta_n(\mathcal{S}) = (X', C')$} (learner.south east);
    
    \node[font=\scriptsize, anchor=south] (program) at ([xshift=-15mm] compiler.west) {Program + Spec.};
    \draw (program.south west) edge[->] (compiler.west);
    \draw (compiler.east) edge[->] node[text width=50mm, align=center, below] {CHC} (teacher.west);
    \node[font=\scriptsize, anchor=north] (safe) at ([yshift=-5mm] teacher.south) {SAFE};
    \draw (teacher.south) edge[->] (safe.north);
    \node[font=\scriptsize, anchor=north] (unsafe) at ([yshift=-5mm] learner.south) {UNSAFE};
    \draw (learner.south) edge[->] (unsafe.north);
  \end{tikzpicture}
  }
  \caption{Graphical summary of the proposed method.} \label{fig:horn-ice-arch}
\end{figure}
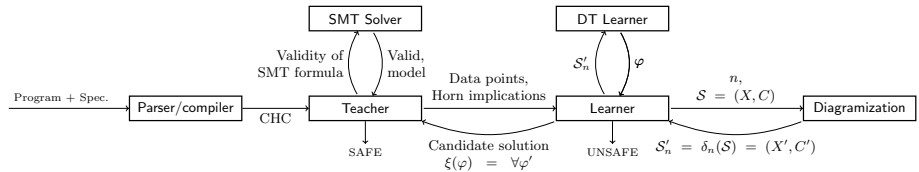
\else
\fi

% and a proof of this is given by and the inconsistency witness, i.e., a data points that need to be classified both as positive (since reachable from the precondition) and negative (since it can reach a state violating the postcondition) provides a proofs that the program is unsafe. 

%Let’s now examine how the number of quantifier variables to be used is determined.
%If the data point sample $\mathcal{S} = (X, C)$ is consistent (there is no data point $x \in X$ that must be classified as both positive and negative based on the constraints in~$C$), but its reduction $\mathcal{S}'_n = (X', C')$ using $n$ quantifier variables is inconsistent (there is a diagram $d \in X'$ such that $x$ is classified as both positive and negative based on the constraints in~$C$), then the chosen number of quantifier variables, $n$, is insufficient as illustrated later in Example~\ref{ex:not-enough-quantifiers}. 
%In such cases, our learner is equipped with an additional loop that increments the number of quantifier variables, starting with one quantifier variable per array, until a consistent diagram sample is obtained.
%This is guaranteed to succeed, as stated in Theorem~\ref{theorem2}, since we can always choose the number of quantifier variables to match the size of the largest array in the sample.
%Notice that if the data point sample~$\mathcal{S}$ is inconsistent then the program is unsafe.

% 2. Programs and Specification
\section{Programs and Specification}
\label{sec:programs-and-spec}

% 2.1 Programs

\ssubsubsection{Programs}
\label{subsec:programs}
In this paper, we consider C-like
programs that manipulate integer-indexed arrays.
Due to space constraints, we only give here an informal description of their
syntax.
Programs contain a designated procedure \texttt{main}
serving as the entry point.
%These programs may involve diverse control flow statements (like conditionals or loops) and use additional procedures, potentially implemented recursively.
%Let $P$ be such a program, and $Proc(P)$ the set of names of procedures of $P$ (including \texttt{main}).
\iflong
  Procedures (except \texttt{main}) can be recursive and may have boolean or integer parameters or pointers to stack-allocated integer-indexed arrays of integers or booleans (see Appendix~\ref{appendix:rec-program} for a recursive program verifiable by our method).
\else
  Procedures (except \texttt{main}) can be recursive and may have boolean or integer parameters or pointers to stack-allocated integer-indexed arrays of integers or booleans.
\fi
In the procedure body, local variables can be declared anywhere.
They can be integer or boolean variables or integer-indexed arrays of integers or booleans; the size of these arrays is parametric and is equal to a linear expression over other integer variables.
We allow various loop structures and conditional statements.
%Let $V_{\Z}$ be the set of integer variables and parameters declared at a particular location.
%In arrays read and write access is only allowed via linear expressions over index variables.
%Loops and conditional statements are constrained to have either linear
%(in)equalities involving index variables or linear (in)equalities of data variables and array reads in their conditions.
%Data variables or arrays' cells can be assigned with linear combinations of data values or array contents.
%Index variables can only be assigned with a linear combination of index variables.\footnote{In the implementation, an automated analysis allows to distinguish the index variables from the data variables by analyzing assignments and conditions. It also detects if the returned value of a procedure must be treated as index or data variables.\label{footnote:data-index-analysis}}

% 2.2 Specification

\ssubsubsection{Specifications}
\label{subsec:spec}
Programs are specified using \texttt{assume}/\texttt{assert} statements
at different program locations.
We introduce here the language of these assumed/asserted properties.
They use the variables of the program at the particular location.

%To each variable we associate a symbol of respective sort (boolean, integer or parametric-size
%integer array).
%If $V$ is the set of variables at the program location of such a statement, we associate to each variable $v \in V$ a symbol~$s$ with the same name. $s$ is a boolean symbol if $v$ is a boolean variable or has the sort $\mathbb{Z}$ if $v$ is an integer variable.
%However, if $v$ is an integer-indexed array variable, $s$ is a parametric-size integer array. \todo{Why having a second
%  set of symbols. That's confusing...}

For expressing the properties as well as the inferred invariants and
pre/post-conditions, we use a many-sorted first-order logic with
one-dimensional arrays
$\BoundedArrayTheory{A}{}$ as
follows. 
The logic $\BoundedArrayTheory{A}{}$ has the following primitive sorts: integers ($Int$) and booleans ($B$) and
two sorts for \emph{finite-size arrays}: integer arrays ($Array(Int)$)
and boolean arrays ($Array(B)$).
Integer constants are $\{\ldots,-1,0,1,\ldots\}$ and
boolean constants are $\{\top,\bot\}$. Integer functions and predicates
are the usual ones of Presburger logic (with the standard
syntactic sugar for linear combinations and comparisons).
Array constants are all finite-size arrays containing either
only integer constants or only boolean constants.
We write them as $[c_0,c_1,\ldots,c_k]$ for some $k \geq 0$ and
$[]$ for the empty array.
Furthermore, we have three functions over integer (boolean) arrays
with the corresponding sorts:
array read $\BoundedArrayRead{\cdot}{\cdot}$,
%of sort $Array(DInt) \times IInt \rightarrow DInt$
%(resp. $Array(B) \times IInt \rightarrow B$),
array write $\BoundedArrayWrite{\cdot}{\cdot}{\cdot}$
%of sort $Array(DInt) \times IInt \times DInt \rightarrow Array(DInt)$
%(resp. $Array(B) \times IInt \times B \rightarrow Array(B)$)
and array length $\BoundedArrayLength{\cdot}$.
For example $a[i]$ is the $i$-th element of array $a$.
%of sort $Array(DInt) \rightarrow IInt$
%(resp. $Array(B) \rightarrow IInt$).
We have also the equality predicate between
two boolean or integer arrays.

Then, \emph{terms}, \emph{atoms},
\emph{literals} and \emph{first-order formulas}
are defined in the usual way.
%Notice that because of the different sorts, a term
%like $a[a[i]]$ is not allowed. Furthermore comparisons between
%index variables and data variables are not allowed.\todo{to be removed ?}
%In the following,
The sort of variables used in the formulas
will be always clear from the context.
The \emph{semantics} can be defined as usual.
Because finite-size arrays are used, we have to
define a semantics for out of bounds access.
Here, instead of using an undefined value, we
just say that for an array read the value of $a[i]$ is $0$ (resp.
$\bot$) for an integer (resp. boolean) array, if $i$ is out of bounds.
%not between $0$ and $\BoundedArrayLength{a}-1$.
An array write $a\{i \leftarrow x\}$ has no effect if $i$ is out of bounds.

Properties in assume/assert statements and verification predicates are \emph{parametric-size array properties}, defined as universally quantified formulas accessing an array~$a$ at indices in the range $0$ to~$\BoundedArrayLength{a} - 1$.

%We restrict the assumed/asserted properties to be well-formed boolean combinations of quantifier-free formulas over the symbols in $S$ and universally quantified parametric-size array properties.
%Any linear combination in the property must be over exclusively either the symbols from $S_{idx}$ or from $S_{data}$.
%The term used as an index in a \textit{read} or \textit{write} operation on an array should not include symbols from $S_{data}$ or any array value (e.g., disallowing expressions like $a[a[i]]$).
%The restriction over the appearence of $S_{idx}$ and $S_{data}$ in the same combination is imposed to avoid falling into an undecidable fragment of the theory.
%Similarly, we constrain parametric-size array properties to adhere to the following form:
\begin{definition}[Parametric-Size Array Property]
  A \emph{parametric-size array property} is a formula of the form
\begin{equation}
\psi \wedge \forall \vec{Q}_{a_1}, \dots ,\vec{Q}_{a_n} .\; \big(\bigwedge_{i=1}^n \bigwedge_{\;\;k \in \vec{Q}_{a_i}} 0 \leq k < \BoundedArrayLength{a_i} \big) \implies \phi(\vec{Q}_{a_1}, \dots ,\vec{Q}_{a_n})
\end{equation}

\noindent
where $\psi$ is a  quantifier free formula of $\BoundedArrayTheory{A}{}$,
the $\vec{Q}_{a_i}$ are index variables and
$\phi(\vec{Q}_{a_1}, \dots ,\vec{Q}_{a_n})$
is a quantifier-free $\BoundedArrayTheory{A}{}$ formula
without array writes in which all read accesses
to arrays $a_i$ are via using one variable of $\vec{Q}_{a_i}$.
\end{definition}

To simplify the presentation we will use formulas
which are syntactically not parametric-size array properties but which are
equivalent to one. %(see example below).
Notice that without further restrictions the satisfiability
of parametric-size array properties is not decidable.
However, one can define a decidable fragment like in \cite{array-theory}
by restricting further $\psi$ and
the use of universally quantified index variables
in $\phi$.

% 2.3 Safety Verification
\ssubsubsection{Safety verification}
\label{subsec:safety-verification}
Given a program and its specification, \textit{safety verification} consists in checking whether along all program executions, whenever all assume statements are satisfied, then also all assert statements are satisfied.
It is well known that this problem amounts to invariant and procedure pre/post-condition synthesis.
In the context of this work, invariants and procedure pre/post-conditions are expressed as universally quantified formulas.
As explained in the overview, we reduce the safety verification of a program to the CHC satisfiability problem.
Then, we define a method for learning a solution of the CHC satisfiability problem in the case of array constraints.
The core of this method is the diagramization technique which is detailed in the following section.

% 3. CHC and CHC Solving
%\input{chc-and-solving}

% 3. Horn-ICE
%\input{horn-ice}

% 4. Learning Quantified Invariants and Summaries using Horn-ICE
%\input{quantified-horn-ice}

% 4. Diagramization
\section{Diagramization}
\label{sec:diagramization}
For an uninterpreted predicate $\Predicate{P} \in \mathcal{P}$, let $\VariablesOf{\Predicate{P}}$ denote the set of all its variable parameters, $\ArraysOf{\Predicate{P}} \subseteq \VariablesOf{\Predicate{P}}$ be the set of its arrays, $\VariablesTypedOf{\Predicate{P}}{\B} \subseteq \VariablesOf{\Predicate{P}}$ the set of its boolean variables and $\VariablesTypedOf{\Predicate{P}}{\Z} \subseteq \VariablesOf{\Predicate{P}}$ the set of its integer variables.  %$V_{\Z}$.

Given a data point sample $\mathcal{S} = (X, C)$, the learner must find
a classifier using quantified formulas over parametric-size arrays 
for $\mathcal{S}$.
We begin with some definitions.
\begin{definition}[Data point Sample] \label{def:datapoint-sample}  
  A data point sample~$\mathcal{S}$ over~$\mathcal{P}$ is a tuple~$(X, C)$, where $X$~is a set of data points over~$\mathcal{P}$, and $C$~is a set of classification constraints over~$X$, represented as Horn implications.
  These constraints take three forms:
  (1) $\top \rightarrow x$, indicating that $x \in X$~must be classified as positive;
  (2) $x_1 \land \cdots \land x_n \rightarrow \bot$, where $x_1, \ldots, x_n \in X$, indicating that at least one of them must be classified as negative;
  (3) $x_1 \land \cdots \land x_n \rightarrow x$, where $x_1, \ldots, x_n, x \in X$, meaning that $x \in X$ must be positive if all $x_1, \ldots, x_n \in X$ are positive;
  Conversely, if $x$~is negative, at least one of $x_1, \ldots, x_n$ must also be negative.
\end{definition}

\begin{definition}[Consistent Labeling of a Data point Sample] \label{def:datapoint-sample-labelling}
  A consistent labeling of a data point sample~$\mathcal{S} = (X, C)$ is a Boolean function~$\mathcal{J} \colon X \to \{\top, \bot\}$ that satisfies all the Horn implications in~$C$.
  Formally, for every implication~$c \in C$, we have
  (1) if $c$~is of the form~$\top \rightarrow x$ (where~$x \in X$), then $\mathcal{J}(x) = \top$;
  (2) if $c$~is of the form~$x_1 \land \cdots \land x_n \rightarrow \bot$ (where~$x_1, \ldots, x_n \in X$), then $\mathcal{J}(x_1) = \bot \lor \cdots \lor \mathcal{J}(x_n) = \bot$;
  (3) if $c$~is of the form~$x_1 \land \cdots \land x_n \rightarrow x$ (where~$x_1, \ldots, x_n, d \in X$), then $\big(\mathcal{J}(x_1) = \top \land \cdots \land \mathcal{J}(x_n) = \top\big)$ implies $\mathcal{J}(x) = \top$.
\end{definition}

\begin{definition}[Consistent Data point Sample]
  A data point sample~$\mathcal{S} = (X, C)$ is said to be \emph{consistent} if there exists a consistent labeling~$\mathcal{J} \colon X \to \{\top, \bot\}$.
  Otherwise, $\mathcal{S}$~is \emph{inconsistent}.
\end{definition}

\begin{definition}[Classifier of a Data point Sample]
  Given a consistent data point sample~$\mathcal{S} = (X, C)$, a \emph{classifier} of~$\mathcal{S}$ is a syntactic characterization of a labeling~$\mathcal{J}$ of~$\mathcal{S}$.
  It is defined as a mapping~$J$ that assigns each predicate~$\Predicate{P} \in \mathcal{P}$ a formula in $\BoundedArrayTheory{A}{}$ over the variables of~$\Predicate{P}$, and that satisfies $x \models J[\PredicateOfDatapoint{x}]$ if and only if~$\mathcal{J}(x) = \top$ for every data point~$x \in X$.
\end{definition}

Our method consists of reducing the classification problem of $\mathcal{S}$ to another classification problem $\mathcal{S}' = (X',C')$
where we do not need quantified formula classifiers.
%We refer to the sample $\mathcal{S}'=(X',C')$ as \textit{diagram sample}.
We will first define the new data points $X'$ and then the
new Horn implications $C'$.
Array values in the data points will be transformed into scalar
values by introducing free variables representing quantifier variables, and scalar variables that take on the values of the array at the positions indicated by the quantifier variables (see Example~\ref{ex:diagram} below).
Then, $C'$ is obtained by modifying $C$.
A classifier for $\mathcal{S}'$ can then be transformed to a classifier for $\mathcal{S}$ by introducing universal quantifiers and substituting the scalar variables with array reads by quantifier variables.

Formally, for each parametric-size array $a \in \ArraysOf{\Predicate{P}}$ of some uninterpreted predicate $\Predicate{P}$, we introduce a set of quantifier variables $\ArrayQuantifierSet{a}{\Predicate{P}}$.
For every $k \in \ArrayQuantifierSet{a}{\Predicate{P}}$, we use a scalar variable $a_{k}$ that has the same type as the elements of the array $a$ and always has the value of $a$ at index $k$.
Let $\ValueSet{\Predicate{P}}$ be the set of these scalar variables, and $\QuantifierSet{\Predicate{P}} = \bigcup_{a \in \ArraysOf{\Predicate{P}}} \ArrayQuantifierSet{a}{\Predicate{P}}$.
We also introduce a fresh integer variable $l_{a}$ representing the size of $a$. Let $\ArrayLengthVariablesOfPredicate{\Predicate{P}}$ be the set of these variables.
In what follows, we define the concept of a diagram.
\begin{definition}[Diagram] \label{def:diagram}
  Let~$x$ be a data point with $\PredicateOfDatapoint{x} = \Predicate{P}$.
  A \emph{diagram}~$d$ of a data point~$x$ is associated with~$\Predicate{P}$ and is a vector over all variables of~$\Predicate{P}$, except the array variables, and $\QuantifierSet{\Predicate{P}} \cup \ValueSet{\Predicate{P}} \cup \ArrayLengthVariablesOfPredicate{\Predicate{P}}$, such that for all variables $v \in \VariablesOf{\Predicate{P}} \setminus \ArraysOf{\Predicate{P}}$, we have $d[v] = x[v]$, and for all arrays $a \in \ArraysOf{\Predicate{P}}$, we have $d[l_{a}] = |x[a]|$ and for all quantifier variables $k \in \ArrayQuantifierSet{a}{\Predicate{P}}$ of $a$, we have $0 \leq d[k] < |x[a]|$, $d[a_{k}] = x[a][k]$.
\end{definition}
We denote with~$\PredicateOfDatapoint{d}$ the uninterpreted predicate associated with diagram~$d$.
Notice that for a data point $x \in X$, there exist multiple diagrams
depending on the size of the arrays of $x$ and the number of introduced
quantifier variables for each array. To simplify, 
%(the cardinality of $\ArrayQuantifierSet{a}{\Predicate{P}}$).
this number is the same for every array.
Let $\Diagrams{n}{x}$ be the set of all diagrams of data point $x$ using
$n$ quantifier variables per array.
\begin{example}\label{ex:diagram}
  For $\ArrayQuantifierSet{a}{\Predicate{I_0}} = \{k_1, k_2\}$, the diagrams of the data point $x_1 = \langle \Predicate{I_0}, N \mapsto 2, a \mapsto [1,0], s \mapsto \bot \rangle$ are $\Diagrams{2}{x_1} = \{d_1, d_2, d_3, d_4\}$ where $d_1, d_2, d_3$ and $d_4$ were introduced in Section~\ref{sec:overview}.
\end{example}
It is possible that different data points have common diagrams:
\begin{example}
The data point~$x_1$ from the previous example and $x_2 = \langle \Predicate{I_0}, N \mapsto 2, a \mapsto [0,0], s \mapsto \bot \rangle$ have the same diagram $\langle \Predicate{I_0}, N \mapsto 2, s \mapsto \bot, k_1 \mapsto 1, k_2 \mapsto 1, a_{k_1} \mapsto 0, a_{k_2} \mapsto 0, l_{a} \mapsto 2 \rangle$ in common.
\end{example}

%The sample $\mathcal{S}'=(X',C')$ obtained from the $\mathcal{S}=(X,S)$ is referred to as a \textit{diagram sample}.
%\begin{definition}
%A \emph{diagram sample} $\mathcal{S'} = (X', C')$ is a set of diagrams $X'$ along with a set of Horn implications $C'$ over the diagrams in $X'$.
%These implications can take one of three forms, similar to those found in a data point sample.
%\end{definition}

We are now ready to define the \emph{diagram sample} $\mathcal{S}'_n=(X',C')$
obtained from $\mathcal{S}=(X,C)$. It is parameterized by the number $n$
of universally quantified variables used.
$X'$ will be the set containing all diagrams
corresponding to data points in~$X$ 
and $C'$ contains Horn implications which bring the constraint
imposed on the classification between data points in~$X$ to the diagrams
in~$X'$.
The intuition is that if all the diagrams associated with a data point~$x$ in~$\mathcal{S}$ are classified positive
by a classifier of~$\mathcal{S}'_n$,
then~$x$ will also be classified positive by a classifier of
$\mathcal{S}$, and conversely.
However, if at least one diagram of a data point~$x$ in~$\mathcal{S}$
is classified negative by a classifier of~$\mathcal{S}'_n$,
then~$x$ will be classified negative by a classifier
of~$\mathcal{S}$ as well, and vice versa.
These constraints are also expressed using Horn implications in~$C'$.

Formally, the notions of diagram sample, along with labeling, consistency, and classifier, are defined analogously to those for data point samples: instead of data points, we have sets of diagrams, and implications are interpreted over diagrams rather than data points.
Note, that for a classifier for a diagram sample, only
formulas over the non-array variables of $\Predicate{P}$ and $\QuantifierSet{\Predicate{P}} \cup \ValueSet{\Predicate{P}} \cup \ArrayLengthVariablesOfPredicate{\Predicate{P}}$ are used.

We can now define the diagram sample constructed from a data point sample.

\begin{definition}
Given a data point sample $\mathcal{S} = (X, C)$ and a parameter~$n$, s.t. $|\ArrayQuantifierSet{a}{\Predicate{P}}| = n$ for all arrays~$a$ in all predicates~$\Predicate{P}$, we obtain a diagram sample using $\delta_n$:
%\begin{equation}
$\delta_n(\mathcal{S}) = \big(\bigcup_{x \in X} \Diagrams{n}{x}, \bigcup_{c \in C} \mu_n(c)\big)$
%\end{equation}
where %$\mu_n(\top \rightarrow x) = \bigcup_{d \in \Diagrams{n}{x}} \{\top \rightarrow d\}$, $\mu_n(x_1 \land \dots \land x_n \rightarrow x_j) = \bigcup_{d_j \in \Diagrams{n}{x_j}} \{\bigwedge_{d \in \bigcup_{x_i \in \{x_1,\dots,x_n\}} \Diagrams{n}{x_i}}d \rightarrow d_j\}$ and $\mu_n(x_1 \land \dots \land x_n \rightarrow \bot) = \{\bigwedge_{d \in \bigcup_{x_i \in \{x_1,\dots,x_n\}} \Diagrams{n}{x_i}} d \rightarrow \bot\}$.
\begin{align*}
\mu_n(\top \rightarrow x) &= \bigcup_{d \in \Diagrams{n}{x}} \{\top \rightarrow d\} \\
\mu_n(x_1 \land \dots \land x_n \rightarrow x_j) &= \bigcup_{d_j \in \Diagrams{n}{x_j}} \{\bigwedge_{d \in \bigcup_{x_i \in \{x_1,\dots,x_n\}} \Diagrams{n}{x_i}}d \rightarrow d_j\} \\
\mu_n(x_1 \land \dots \land x_n \rightarrow \bot) &= \{\bigwedge_{d \in \bigcup_{x_i \in \{x_1,\dots,x_n\}} \Diagrams{n}{x_i}} d \rightarrow \bot\}
\end{align*}
\end{definition}
\begin{example}
For instance, Fig.\ref{fig:diagram-sample} shows the diagram sample derived from the data point sample in Fig.\ref{fig:data-sample}.
\end{example}

Notice that even if the CHC system is linear, the obtained Horn implications will be nonlinear because of the presence of arrays.
%Notice that even if the CHC system is linear, and thus the implications are linear in the data point sample~$S$, through diagramization, we will obtain a diagram sample with Horn implications.
%A classifier for~$\mathcal{S}'_n$ gives for each predicate $\Predicate{P} \in \mathcal{P}$ a quantifier-free formula over non-array variables of $\Predicate{P}$ and $\QuantifierSet{\Predicate{P}} \cup \ValueSet{\Predicate{P}} \cup \ArrayLengthVariablesOfPredicate{\Predicate{P}}$.

%In our case, we define a solution for a diagram sample as follows.\todo{is this definition necessary ?}
%\begin{definition}
%A \emph{solution} $J'$ of a diagram sample $\mathcal{S}' = (X', C')$ maps uninterpreted predicates $\Predicate{P} \in \mathcal{P}$ to quantifier-free formulas over non-array variables of $\Predicate{P}$ and $\QuantifierSet{\Predicate{P}} \cup \ValueSet{\Predicate{P}} \cup \ArrayLengthVariablesOfPredicate{\Predicate{P}}$.
%These predicates respect the implications in $C'$.
%\end{definition}

Given a classifier $J'$ for the sample $\mathcal{S}'_n := \delta_n(\mathcal S)$ we can
construct a classifier~$J$ for the data point sample~$\mathcal{S}$ by quantifying the introduced quantifier variables, substituting the scalar variables with reads of the arrays with their respective quantifier variables (i.e., substituting~$a_k$ with~$a[k]$), and replacing the introduced size variables with the sizes of their respective arrays (i.e., substituting~$l_a$ with~$|a|$). Formally,
\begin{definition}
Let~$J'$ be a classifier for~$\delta_n(\mathcal{S})$, we define~$\xi$ such that for every uninterpreted predicate $\Predicate{P} \in \mathcal{P}$, 
%\begin{align*}
$\xi(J')[\Predicate{P}] = \forall \vec{Q}_{a_1}, \dots ,\vec{Q}_{a_n} .\; \big(\bigwedge_{i=1}^n \bigwedge_{\;\;k \in \vec{Q}_{a_i}} 0 \leq k < \BoundedArrayLength{a_i} \big) \implies J'[\Predicate{P}][a_k/a[k],l_a/|a|]_{a \in \ArraysOf{\Predicate{P}}}$
%\end{align*}
where the substitution is for all arrays of $\Predicate{P}$.
\end{definition}

We have the following theorem showing that the construction is correct
allowing to obtain a classifier of the data point sample from a classifier
of the diagram sample.

\begin{figure}[ht]
  
\centering
\begin{subfigure}[b]{1\textwidth}
\centering
\scalebox{0.7}{
  \begin{tikzpicture}
    \node[anchor=east] (0) at (-3.5, -0.25) {$\top$};

    \node[anchor=east] (11) at (0, 0) {$\langle \Predicate{I_0}, 2, [0, 0], \top \rangle$};
    \node[anchor=east] (12) at (3.5, 0) {$\langle \Predicate{I_1}, 2, [0, 0], 1, \bot \rangle$};
    \node[anchor=east] (13) at (7, 0) {$\langle \Predicate{I_1}, 2, [0, 0], 2, \bot \rangle$};
    \node[anchor=east] (14) at (10.5, 0) {$\langle \Predicate{I_0}, 2, [0, 0], \bot \rangle$};

    \node[anchor=east] (21) at (0, -0.5) {$\langle \Predicate{I_0}, 2, [1, 1], \top \rangle$};
    \node[anchor=east] (22) at (3.5, -0.5) {$\langle \Predicate{I_1}, 2, [1, 1], 1, \bot \rangle$};
    \node[anchor=east] (23) at (7, -0.5) {$\langle \Predicate{I_1}, 2, [1, 1], 2, \bot \rangle$};
    \node[anchor=east] (24) at (10.5, -0.5) {$\langle \Predicate{I_0}, 2, [1, 1], \bot \rangle$};

    \node[anchor=east] (31) at (10.5, -1) {$\langle \Predicate{I_0}, 2, [1, 0], \bot \rangle$};

    \node[anchor=east] (1) at (12, -1) {$\bot$};

    \draw[->, tips=on proper draw] (0.east) edge (11.west);
    \draw[->, tips=on proper draw] (11.east) edge (12.west);
    \draw[->, tips=on proper draw] (12.east) edge (13.west);
    \draw[->, tips=on proper draw] (13.east) edge (14.west);

    \draw[->, tips=on proper draw] (0.east) edge (21.west);
    \draw[->, tips=on proper draw] (21.east) edge (22.west);
    \draw[->, tips=on proper draw] (22.east) edge (23.west);
    \draw[->, tips=on proper draw] (23.east) edge (24.west);

    \draw[->, tips=on proper draw] (31.east) edge (1.west);
  \end{tikzpicture}
}
\setlength{\abovecaptionskip}{-2ex}
   \caption{}
   \label{fig:ex:quantifier-not-enough-1} 
\end{subfigure}

\begin{subfigure}[b]{1\textwidth}
\centering
\scalebox{0.7}{
  \begin{tikzpicture}
    %\draw[color=gray] (-4,-4) grid (14,4);

    \node[anchor=east] (0) at (-3.5, -0.75) {$\top$};

    \node[anchor=east] (111) at (0, 0) {$\langle \Predicate{I_0}, 2, \top, 0, 0, 2 \rangle$};
    \node[anchor=east] (112) at (0, -0.5) {$\langle \Predicate{I_0}, 2, \top, 1, 0, 2 \rangle$};
    \node[anchor=east] (121) at (3.5, 0) {$\langle \Predicate{I_1}, 2, 1, \bot, 0, 0, 2 \rangle$};
    \node[anchor=east] (122) at (3.5, -0.5) {$\langle \Predicate{I_1}, 2, 1, \bot, 1, 0, 2 \rangle$};
    \node[anchor=east] (131) at (7, 0) {$\langle \Predicate{I_1}, 2, 2, \bot, 0, 0, 2 \rangle$};
    \node[anchor=east] (132) at (7, -0.5) {$\langle \Predicate{I_1}, 2, 2, \bot, 1, 0, 2 \rangle$};
    \node[anchor=east] (141) at (10.5, 0) {$\langle \Predicate{I_0}, 2, \bot, 0, 0, 2 \rangle$};
    \node[anchor=east] (142) at (10.5, -0.5) {$\langle \Predicate{I_0}, 2, \bot, 1, 0, 2 \rangle$};

    \node[anchor=east] (211) at (0, -1) {$\langle \Predicate{I_0}, 2, \top, 0, 1, 2 \rangle$};
    \node[anchor=east] (212) at (0, -1.5) {$\langle \Predicate{I_0}, 2, \top, 1, 1, 2 \rangle$};
    \node[anchor=east] (221) at (3.5, -1) {$\langle \Predicate{I_1}, 2, 1, \bot, 0, 1, 2 \rangle$};
    \node[anchor=east] (222) at (3.5, -1.5) {$\langle \Predicate{I_1}, 2, 1, \bot, 1, 1, 2 \rangle$};
    \node[anchor=east] (231) at (7, -1) {$\langle \Predicate{I_1}, 2, 2, \bot, 0, 1, 2 \rangle$};
    \node[anchor=east] (232) at (7, -1.5) {$\langle \Predicate{I_1}, 2, 2, \bot, 1, 1, 2 \rangle$};
    \node[anchor=east] (241) at (10.5, -1) {$\langle \Predicate{I_0}, 2, \bot, 0, 1, 2 \rangle$};
    \node[anchor=east] (242) at (10.5, -1.5) {$\langle \Predicate{I_0}, 2, \bot, 1, 1, 2 \rangle$};

    \node[anchor=east] (1) at (12, -0.75) {$\top$};

    \draw[->, tips=on proper draw] (0.east) edge (111.west);
    \draw[->, tips=on proper draw] (0.east) edge (112.west);
    \draw[->, tips=on proper draw] (0.east) edge (211.west);
    \draw[->, tips=on proper draw] (0.east) edge (212.west);

    \draw[->] (111.east) -- ++(0.3, 0) coordinate (121in) -- (121.west);
    \draw[-] (112.east) -- (121in);
    \draw[fill=white] (121in) circle (1.1pt);

    \draw[->] (121.east) -- ++(0.3, 0) coordinate (131in) -- (131.west);
    \draw[-] (122.east) -- (131in);
    \draw[fill=white] (131in) circle (1.1pt);

    \draw[->] (131.east) -- ++(0.5, 0) coordinate (141in) -- (141.west);
    \draw[-] (132.east) -- (141in);
    \draw[fill=white] (141in) circle (1.1pt);

    \draw[->] (112.east) -- ++(0.3, 0) coordinate (122in) -- (122.west);
    \draw[-] (111.east) -- (122in);
    \draw[fill=white] (122in) circle (1.1pt);

    \draw[->] (122.east) -- ++(0.3, 0) coordinate (132in) -- (132.west);
    \draw[-] (121.east) -- (132in);
    \draw[fill=white] (132in) circle (1.1pt);

    \draw[->] (132.east) -- ++(0.5, 0) coordinate (142in) -- (142.west);
    \draw[-] (131.east) -- (142in);
    \draw[fill=white] (142in) circle (1.1pt);

    \draw[->] (211.east) -- ++(0.3, 0) coordinate (221in) -- (221.west);
    \draw[-] (212.east) -- (221in);
    \draw[fill=white] (221in) circle (1.1pt);

    \draw[->] (221.east) -- ++(0.3, 0) coordinate (231in) -- (231.west);
    \draw[-] (222.east) -- (231in);
    \draw[fill=white] (231in) circle (1.1pt);

    \draw[->] (231.east) -- ++(0.5, 0) coordinate (241in) -- (241.west);
    \draw[-] (232.east) -- (241in);
    \draw[fill=white] (241in) circle (1.1pt);

    \draw[->] (212.east) -- ++(0.3, 0) coordinate (222in) -- (222.west);
    \draw[-] (211.east) -- (222in);
    \draw[fill=white] (222in) circle (1.1pt);

    \draw[->] (222.east) -- ++(0.3, 0) coordinate (232in) -- (232.west);
    \draw[-] (221.east) -- (232in);
    \draw[fill=white] (232in) circle (1.1pt);

    \draw[->] (232.east) -- ++(0.5, 0) coordinate (242in) -- (242.west);
    \draw[-] (231.east) -- (242in);
    \draw[fill=white] (242in) circle (1.1pt);

    \draw[<-] (1.west) -- ++(-0.5, 0) coordinate (1in) -- (241.east);
    \draw[-] (142.east) -- (1in);
    \draw[fill=white] (1in) circle (1.1pt);

  \end{tikzpicture}
}\setlength{\abovecaptionskip}{-0.5ex}
   \caption{}
   \label{fig:ex:quantifier-not-enough-2}
\end{subfigure}
\setlength{\abovecaptionskip}{0ex}
\setlength{\belowcaptionskip}{-3ex}
\caption{(a) A data point sample which has a classifier and (b) its (non consistent) diagram sample using only one quantifier per array does not admit any classifier.}
\label{fig:ex:quantifier-not-enough}
\end{figure}
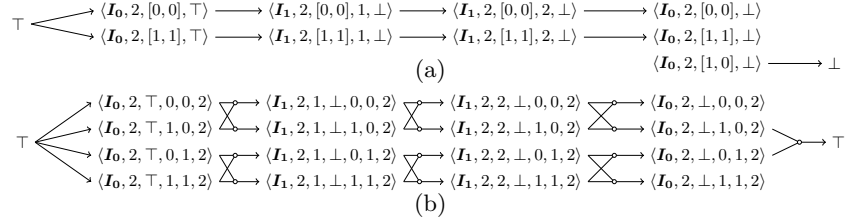

\begin{theorem}\label{theorem1}
  Let $\mathcal{S} = (X, C)$ be a consistent data point sample and let
  $\mathcal{S}'_n = \delta_n(\mathcal{S})$ be the corresponding diagram sample.
  If $\mathcal{S}'_n$ is consistent and $J'$ is a classifier of $\mathcal{S}'_n$ then~$\xi(J')$ is a classifier of $\mathcal{S}$.
\end{theorem}
\iflong
All the proofs are deferred to the Appendix~\ref{appendix:proofs}.
\else
All the proofs are deferred to the full paper~\cite{cav25arxiv}.
\fi
Notice that the existence of a classifier for~$\mathcal{S}'_n$ depends on the number $n$ of quantifier variables introduced per array.
It is possible that  %n abstracted
sample~$\mathcal{S}'_n$ has no classifier (because it is inconsistent),
despite the existence of a
classifier for~$\mathcal{S}$, e.g. in the following.
\begin{example}\label{ex:not-enough-quantifiers}
  Fig.~\ref{fig:ex:quantifier-not-enough-1} shows a data point sample~$\mathcal{S}$, of the program in
  Fig.~\ref{fig:bubble-sort}. It has a classifier but its corresponding diagram sample using only one quantifier per array (Fig.\ref{fig:ex:quantifier-not-enough-2}) has no classifier (since it is inconsistent) as both $\langle \Predicate{I_0}, 2, \bot, 1, 0, 2 \rangle$ and $\langle \Predicate{I_0}, 2, \bot, 0, 1, 2 \rangle$ read as $\langle \Predicate{I_0}, N \mapsto 2, s \mapsto \bot, k_1 \mapsto 0, a_{k_1} \mapsto 1, l_{a} \mapsto 2 \rangle$, are forced to be satisfied by
  $\Predicate{I_0}$ while the implication that connects them to~$\bot$ requires that at least one of them must not (In~\ref{fig:ex:quantifier-not-enough-2}, $\langle \Predicate{I_1}, 2, 1, \bot, 0, 0, 2 \rangle$ is read as $\langle \Predicate{I_1}, N \mapsto 2, i \mapsto 1, s \mapsto \bot, k_1 \mapsto 0, a_{k_1} \mapsto 0, l_{a} \mapsto 2 \rangle$).
\end{example}

In such a situation, we increase $n$ until the diagram sample is consistent, as the following theorem shows that there exists a sufficient number of quantifiers per array for which~$\mathcal{S}'_n$ is consistent if~$\mathcal{S}$ is.

\begin{theorem}\label{theorem2}
  Let $\mathcal{S} = (X, C)$ be a consistent data point sample. If, for every predicate~$\Predicate{P}$ and for every array~$a$ in the domain of~$\Predicate{P}$, we have $|\ArrayQuantifierSet{a}{\Predicate{P}}| \geq |d[a]|$ for every diagram $d \in \Diagrams{n}{x}$ of every data point $x \in X$, then
  $\mathcal{S}'_n = \delta_n({\mathcal S})$ is also consistent.
\end{theorem}

The diagram sample size grows exponentially with respect to the number of introduced quantifier variables.
This potential combinatorial explosion can be mitigated by observing that the number of diagrams in the diagram sample can be reduced by imposing a particular order on the quantifier variables of the same array $k_1, \dots, k_n$, e.g. in example~\ref{ex:diagram} we remove the third diagram.
Therefore, the index guard of the property constructed using $\xi$ is conjuncted with $k_1 \leq \dots \leq k_n$ for the quantifier variables $k_1, \dots, k_n$ of every array $a$.
This is justified by $\forall k_1, k_2.\; \varphi(k_1,k_2)$ being equivalent to $\forall k_1,k_2.\; k_1 \leq k_2 \implies \varphi(k_1,k_2) \land \varphi(k_2,k_1)$ (can be generalized to more than $2$ quantifiers).
The diagram sample also depends on the size of the arrays in the data points, with a preference for arrays of smaller size.
To take advantage of this, an optimization is applied by tuning the teacher to produce counterexamples with smaller arrays.

% 5. QDT
\section{Decision Tree-based Quantified Invariants Learner}
\label{sec:qdt-learner}

Here, we present the algorithm of a decision-tree-based learner for synthesizing universally quantified properties
using the diagramization primitives introduced earlier
and explain how the attributes fed into the decision-tree learning algorithm are constructed.

\ssubsubsection{The learner}
We provide an instantiation of the learner of the Horn-ICE framework capable of synthesizing universally quantified properties in Fig.~\ref{fig-algo:qdt-learner}.
The learner maintains a variable~$n$ representing the number of quantifiers to be used per array (line~\ref{qdt:line:n}) and $Attributes$ which
maps predicates to a set of attributes (initialized in line~\ref{qdt:line:attributes}).
In each iteration, the learner is invoked with a given data point sample~$\mathcal{S}$
and learns a solution for it starting from the current state parameters ($n$~and~$Attributes$) by constructing a diagram sample~$\mathcal{S}'_n$ from~$\mathcal{S}$ with $n$ quantifiers per array (line~\ref{qdt:line:construct-sp}).
If this diagram sample is inconsistent, $n$ is incremented until a constructed sample is consistent (loop in lines~\ref{qdt:line:sample-consistent}-\ref{qdt:line:construct-sp-2}).
Note that this consistency check can be performed in polynomial time since the classification constraints are expressed as Horn implications.
Then, the learner checks if the attributes are sufficient to classify the sample~$\mathcal{S}'$.
This is detected by calling $\textsc{Sufficient}$ (line~\ref{qdt:line:sufficient}).
If this is not the case, more attributes are generated (line~\ref{qdt:line:generate-attributes}) until they are deemed sufficient (loop in lines~\ref{qdt:line:sufficient}-\ref{qdt:line:generate-attributes}).
Once the diagram sample is consistent and the attributes are known to be sufficient, it learns a quantifier-free solution for $\mathcal{S}'$ using $\textsc{Decision-Tree-Horn}$ (line~\ref{qdt:line:learn-dt}).
Once the solution~$J'$ for~$\mathcal{S}'_n$ is found, a solution~$J$ for~$\mathcal{S}$ is then constructed from it and returned (line~\ref{qdt:line:return}).

\begin{figure*}[ht]
  \centering
  %\footnotesize
  \begin{minipage}{.79\linewidth}%\scriptsize
    \begin{algorithm}[H]
      \SetKwInOut{Input}{Input}
      \SetKwInOut{Output}{Output}
      \SetKwProg{Proc}{Proc}{}{}
    
      \Input{A data point sample $\mathcal{S}$ over predicates $\mathcal{P}$}
      \Output{A candidate solution $J$ of $\mathcal{S}$}
      $n \gets \text{Initial number of quantifiers per array}$\;\label{qdt:line:n}
      $Attributes \gets \text{Initial set of attributes}$\;\label{qdt:line:attributes}
      \Proc{$\textsc{Learner.learn}(\mathcal{S})$}{
        $\mathcal{S'} \gets \delta_{n}(\mathcal{S})$\;\label{qdt:line:construct-sp}
        \While{$\neg \textsc{consistent}(S')$}{\label{qdt:line:sample-consistent}
          $n \gets n + 1$; \label{qdt:line:incr-n}
          $\mathcal{S'} \gets \delta_{n}(\mathcal{S})$\label{qdt:line:construct-sp-2}
        }
        \While{$\neg \textsc{sufficient}(Attributes,S')$}{\label{qdt:line:sufficient}
          $Attributes \gets \textsc{generateAttributes}(Attributes,S')$\;\label{qdt:line:generate-attributes}
        }
        $J' \gets \textsc{Decision-Tree-Horn}(S', Attributes)$\;\label{qdt:line:learn-dt}
        \KwRet{$\xi(J')$}\;\label{qdt:line:return}
      }
    \end{algorithm}
  \end{minipage}%\setlength{\abovecaptionskip}{-1ex}
\setlength{\belowcaptionskip}{-4ex}
  \caption{The quantified interpretations learner.}\label{fig-algo:qdt-learner}
\end{figure*}
$\textsc{Decision-Tree-Horn}$ is straightforwardly adapted from Horn-ICE-DT~\cite{horn-ice} and applied on a diagram sample instead of a data point sample.
If the attributes are sufficient, it constructs
a quantifier-free formula for each predicate~$\Predicate{P}$
by combining attributes in~$Attributes[\Predicate{P}]$ using the decision-tree learning algorithm.
$\textsc{Sufficient}$ checks if the attributes are sufficient to construct classifying decisions trees for the sample by computing equivalence classes, as described in~\cite{ice-dt}.

\ssubsubsection{Attribute Discovery}
In our approach, attributes for an uninterpreted predicate $\Predicate{P}$ are defined as atomic formulas over scalar variables associated with $\Predicate{P}$.
These variables include non-array variables appearing in the diagrams of $\Predicate{P}$ and variables drawn from the set $\QuantifierSet{\Predicate{P}} \cup \ValueSet{\Predicate{P}} \cup \ArrayLengthVariablesOfPredicate{\Predicate{P}}$.
The attribute set for each predicate $\Predicate{P}$ is maintained by the mapping $Attributes$ (line~\ref{qdt:line:attributes}).
This set is constructed using a finite collection of attribute patterns, which fall into two broad categories:
\begin{itemize}
  \item Enumerated Patterns: These are syntactically defined templates (e.g., arithmetic constraints) that are instantiated using all possible combinations of the relevant scalar variables.
  Some patterns involve constants and are enumerated incrementally by increasing a bound $k$ on the absolute values of constants.
  The types of constraints considered here include namely intervals ($\pm v \le c$), upper bounds ($v_1 \le v_2$), or octagons ($\pm v_1 \pm v_2 \le c$) where $v$, $v_1$, $v_2$ range over the relevant scalar variables and $c \in \Z$ with $|c| \le k$.
  \item Extracted Patterns: These are patterns derived from the program itself, particularly from conditional and assignment statements and from specification constructs such as assume and assert statements.
  For example, from a program assignment like \lstinline[columns=fixed,breaklines=true]{c[i] = a[i] - b[i];}, we extract the pattern $v_1 = v_2 - v_3$, and instantiate it over the variable set $\VariablesTypedOf{\Predicate{P}}{\Z} \cup \QuantifierSet{\Predicate{P}} \cup \ValueSet{\Predicate{P}} \cup \ArrayLengthVariablesOfPredicate{\Predicate{P}}$.
\end{itemize}
This dual strategy balances expressiveness and scalability: we restrict enumeration to tractable forms (interval and octagonal), while allowing more complex, potentially nonlinear constraints to be captured through pattern extraction from program logic.
When the current attribute set in $Attributes$ is insufficient to classify the sample of diagrams, the function $\textsc{generateAttributes}$ is invoked.
This function increases the constant bound~$k$ and re-instantiates the enumerated patterns with the extended constant range.
Because the attribute space strictly increases with~$k$, it follows that for a sufficiently large~$k$, any pair of distinct diagrams can eventually be separated by an appropriate attribute.
%This ensures completeness of the method in the limit.

% 6. Experiments
\section{Experiments}

We have implemented our method in the tool \textsc{Tapis}\footnote{An artifact that includes \textsc{Tapis} and all the benchmarks is available online~\cite{tapis}.} (Tool for Array Program Invariant Synthesis) written in C++.
It uses Clang%\footnote{\url{https://clang.llvm.org}}
as the frontend for parsing/type-checking admissible C programs, as well as Z3~\cite{z3} for checking satisfiability of SMT queries.
Given a program, \textsc{Tapis} generates the verification conditions of the program as CHCs with parametric-size arrays from its type-annotated AST.
The CHC satisfiability problem is fed to the learning loop using our learner described in Section~\ref{sec:qdt-learner} and Z3 as the teacher verifying the validity of the proposed solutions produced by the learner and translated to SMT queries.
The teacher is tuned to find counterexamples with small array sizes.
It discharges SMT queries to identify counterexamples with arrays bounded by~$L$ (initially set to~$1$).
If no counterexample is found, the bound is removed for another check.
At this point, either the formula is valid, or $L$~is incremented, and the process repeats.
Without this tuning, Z3 tends to generate counterexamples with excessively large array sizes (e.g., \textgreater 1000), leading to large diagram samples during diagramization, which can cause timeouts.

We compare \textsc{Tapis} with \textsc{Spacer}, \textsc{Ultimate Automizer}, \textsc{FreqHorn}, \textsc{Vajra}, \textsc{Diffy}, \textsc{Rapid}, \textsc{Prophic3} and \textsc{MonoCera}.
\textsc{Spacer}~\cite{spacer} is a PDR-based CHC solver integrated in Z3
using \textsc{Quic3}~\cite{quic3} for universal quantifier support.
\textsc{UAutomizer}~\cite{uautomizer} is a program verification tool combining counterexample-guided abstraction refinement with trace abstraction.
\textsc{FreqHorn}~\cite{freqhorn,freqhorn-quantifiers} is a syntax-guided synthesis CHC solver extended to synthesize quantified properties.
\textsc{Vajra}~\cite{vajra} implements a full-program induction technique to prove quantified properties of parametric size array-manipulating programs.
\textsc{Diffy}~\cite{diffy} improves \textsc{Vajra}'s full-program induction with difference invariants.
\textsc{Rapid}~\cite{rapid} is a verification tool for array programs, specialized in inferring invariants with quantifier alternation using trace logic.
\textsc{Prophic3}~\cite{prophic3} employs counterexample-guided abstraction refinement with prophecy variables to reduce array program verification to quantifier-free and array-free reasoning.
\textsc{Prophic3} is built on top of~\textsc{ic3ia}~\cite{ic3ia}.
\textsc{MonoCera}~\cite{monocera} implements an instrumentation-based method and handles specifications involving aggregation and quantification.
\textsc{MonoCera} is built on top of~\textsc{TriCera}~\cite{tricera}.
\textsc{Spacer}, \textsc{FreqHorn}, \textsc{Vajra}, \textsc{Diffy}, \textsc{Rapid}, \textsc{Prophic3} and \textsc{Tapis} are written in C++, while \textsc{UAutomizer} is written in Java and \textsc{MonoCera} is written in Scala.

\textsc{Tapis}, \textsc{UAutomizer}, \textsc{Vajra}, \textsc{Diffy} and \textsc{MonoCera} support different subsets of C programs with parametric arrays.
\textsc{Rapid} accepts programs in its own custom language.
For our experimental comparisons, benchmark programs are manually translated into the language or program class accepted by each tool.
\textsc{Spacer} and \textsc{FreqHorn} require CHC problems in the SMT-LIB format~\cite{smt-lib}.
After generating the verification conditions of the program as CHCs, our tool exports them in SMT-LIB format for use with \textsc{Spacer} and \textsc{FreqHorn}.
\textsc{Prophic3} takes symbolic transition systems in the VMT format~\cite{vmt}.
We use \textsc{Kratos2}~\cite{kratos2} (and \texttt{c2kratos.py}) to convert C programs into the corresponding transition systems in the VMT format.

We compare the tools on two benchmark sets of 215~programs:
The first set consists of all array programs from SV-COMP\footnote{All C programs in the \texttt{c/array-*} directories of SV-Benchmarks~\url{https://gitlab.com/sosy-lab/benchmarking/sv-benchmarks}.}~\cite{sv-comp}
except those involving dynamic memory, pointer arithmetic, and/or aggregations.
These programs are modified by replacing fixed array sizes by parametric sizes.
They are categorized into 84 safe programs and 37 unsafe ones.
The first benchmark set (1) does not include recursive programs and (2) primarily consists of programs with simple iteration patterns, such as \lstinline[columns=fixed,breaklines=true]{for(i=...; i<...; i++)}.
Therefore, we have constructed a second benchmark set that includes the lacking types of programs from the first set.
This second benchmark, which we call \emph{TAPIS-Bench}, includes sorting algorithms—specifically, insertion sort and quicksort, which are absent from the SV-COMP benchmark—as well as versions of SV-COMP sorting algorithms without partial annotations.
Additionally, it icludes other array algorithms with diverse looping patterns, such as \lstinline[columns=fixed,breaklines=true]{for(i=...; i<...; i++)}, \lstinline[columns=fixed,breaklines=true]{for(j=...; j>...; j--)}, and \lstinline[columns=fixed,breaklines=true]{for(i=..., j=...; j>i; i++, j--)}.
These algorithms are implemented in both iterative and recursive versions.
They are all safe and they are categorized into 49 \emph{iterative} non-procedural programs, 37 (\emph{rec}) (non-mutually) recursive procedures and 8 (\emph{mut-rec}) with mutually recursive procedures.
These programs are not partially annotated, and proving their safety requires synthesizing an invariant for each loop and a pre/post-condition for every procedure (except \texttt{main}).
Programs with tail recursive procedures in the \emph{rec} category have their iterative equivalents in the \emph{iterative} category.
Across the two benchmark sets, the number of procedures (including \texttt{main}) ranges from 1 to 3, while the number of loops varies between 1 and 9.
%\iflong
%Appendix~\ref{appendix:benchmarks} contains a detailed enumeration of all programs featured in the benchmark.
%\else
%In the supplementary material, the Appendix~C of the extended version of this paper contains a detailed enumeration of all programs featured in the benchmark.
%\fi
The evaluation was carried out using a timeout of 300s for each example on 
an 8 cores 3.2GHz CPU with 16 Go RAM.

We do not consider the CHC-COMP~\cite{chc-comp} benchmark set as they are over unbounded arrays and incompatible with our method because we require fixed-size array values in the counterexamples.

\begin{table}
  \centering
  \caption{Benchmark results.}
  \label{tab:benchmarks}
  \begin{tabular}{c||c|c|c|c|c||c|c|c}
    \hline
    \multirow{3}{*}{Tool} & \multicolumn{2}{c|}{SV-COMP} & \multicolumn{3}{c||}{TAPIS-Bench} & \multicolumn{3}{c}{Total} \\
    \cline{2-9}
                      & \multirow{2}{*}{\makecell{safe \\ (84)}} & \multirow{2}{*}{\makecell{unsafe \\ (37)}} & \multirow{2}{*}{\makecell{iterative \\ (49)}} & \multirow{2}{*}{\makecell{rec \\ (37)}} & \multirow{2}{*}{\makecell{mut-rec \\ (8)}} & \multicolumn{2}{c|}{safe (178)} & \multirow{2}{*}{\makecell{all \\ (215)}} \\\cline{7-8}
                      & & & & & & \makecell{iterative \\ (133)} & \makecell{recursive \\ (45)} \\
    \hline
    \textsc{Tapis}     &  48 &     19 &        {\bf 47} &  {\bf 37} &       {\bf 8} & {\bf 95} & {\bf 45} &   {\bf 159} \\
    \textsc{Spacer}    &  50 &     30 &        37 &  14 &       1 & 87 & 15 &   132 \\
    \textsc{Prophic3}  &  {\bf 56} &     32 &        30 &   - &       - & 86 & - &   118 \\
    \textsc{MonoCera}  &  25 &     20 &        24 &  12 &   0 &      49 &  12 & 81 \\
    \textsc{Diffy}     &  41 &     23 &         9 &   - &       - & 50 & - &    73 \\
    \textsc{Vajra}     &  42 &     23 &         8 &   - &       - & 50 & - &   73 \\
    \textsc{UAutomizer}&  12 &     {\bf 34} &         7 &   4 &       1 & 19 & 5 &  58 \\
    \textsc{FreqHorn}  &  44 &      2 &        11 &   - &       - & 55 & - &  57 \\
    \textsc{Rapid}     &   3 &      - &         3 &   - &       - & 3 & - &   6 \\
    \hline
  \end{tabular}
\end{table}
The results are shown in Table~\ref{tab:benchmarks}.
The total columns aggregate results from SV-COMP and TAPIS-Bench.
We did not include the total count of unsafe programs, as they correspond to the SV-COMP/unsafe column.
The results show that overall, within the fixed timeout, among the 215~programs, \textsc{Tapis} successfully solves 159 programs, surpassing \textsc{Spacer} by 27 programs and \textsc{Prophic3} by 41 programs.
\textsc{Tapis}, despite not being specialized in proving unsafety, successfully solves 19 unsafe programs.
The effectiveness of \textsc{Spacer} is based on its capacity to generalize the set of predecessors computed using model-based projection and interpolants.
This task becomes particularly challenging in the presence of quantifiers, especially when dealing with non-linear CHC or those containing multiple uninterpreted predicates to infer.
This is notably evident in the context of solving recursive programs.
\textsc{Tapis} solves 8 more iterative safe programs and 30 more recursive programs than \textsc{Spacer}.
The reduction of array programs to a quantifier-free array-free problem enables \textsc{Prophic3} to solve a significant number of safe programs.
Moreover, its foundation on \textsc{ic3ia}, a PDR/IC3-based approach, enhances its effectiveness in solving unsafe programs, similar to \textsc{Spacer}, which is also built on PDR/IC3.
Differently from \textsc{Spacer}, \textsc{Prophic3} cannot solve programs with recursion.
\textsc{MonoCera} has successfully verified a total of 81 programs, including 12 recursive ones.
Its effectiveness, however, strongly depends on the predefined instrumentation schema used for universal quantification.
Consequently, its verification capabilities are largely confined to relatively simple array traversals, especially those in which the necessary invariants are closely aligned with the properties to be verified.
Many safe programs from SV-COMP fall within the restrictive class accepted by \textsc{Vajra} and \textsc{Diffy}.
However, these tools are limited in handling programs with different looping patterns from the \emph{iterative} category of TAPIS-Bench.
\textsc{UAutomizer} is based on a model-checking approach and, although it can effectively solve instances with fixed-size arrays, it faces challenges in the parametric case.
However, it is effective in verifying unsafe programs, solving the highest number of such cases.
\textsc{FreqHorn}, on the other hand, only manages to solve 57 programs overall.
It can not verify recursive programs as it only supports linear CHC.
Additionally, \textsc{FreqHorn} encounters issues when handling programs with quantified preconditions/assumptions which it can not handle.
The failure of \textsc{FreqHorn} to solve many other programs from the \emph{iterative} category can be attributed to its range analysis.
Notably, \textsc{FreqHorn} struggles in solving identical algorithms when implemented with different iterating patterns (like iterating from the end of the array or using two iterators simultaneously from both the beginning and end).
\textsc{Rapid}, which specializes in inferring invariants with quantifier alternation, successfully solves only 6 programs, limited by its custom language's inability to represent programs with procedure calls or quantified preconditions.
The tool's effectiveness hinges on the capability of its customized \textsc{Vampire}~\cite{vampire} theorem prover to tackle the generated reasoning tasks, which pose significant challenges.

Handling parametric size arrays in program verification significantly enhances scalability compared to tools limited to fixed-size arrays.
For instance,
\textsc{UAutomizer} solves the program \texttt{array-argmax-fwd} from the TAPIS-Bench with the array-size parameter~$N$ set to 2 in 83 seconds, and 191 seconds for~$N$ set to 4.
However, the resolution time surpasses 36 minutes when~$N$ is increased to 5.
Conversely, \textsc{Tapis} demonstrates its efficiency by solving the same program in just 0.47 seconds for arbitrary~$N$.

The benchmark sets include programs where the number of data points exceeds 300, with more than 1200 diagrams in the last iteration.
These numbers can grow even larger for programs where our method timeouts.
However, we do not report them explicitly, as they vary across executions due to Z3's non-determinism.
\textsc{Tapis} never exceeds two quantifier variables per array per predicate in the two benchmark sets.
While some invariants may require only a single quantifier variable, our method occasionally necessitates more, as each quantifier variable can only be used to access a single array.

%We notice that the use of the optimisations mentioned in the beginning of the section has a significant impact.
%For instance, the use of the light-weight analysis to discover basic equalities between program variables before the learning loop allows to solves 27 more programs compared to when it is disabled.

\pgfplotsset{runtime diagram/.style={
  xmin = 1e-2, xmax = 700,
  ymin = 1e-2, ymax = 700,
  log basis x=10,
  log basis y=10,
  enlarge x limits=false, enlarge y limits=false,
  xtickten={-1,...,1},
  ytickten={-1,...,1},
  extra x ticks={3e2,700}, extra x tick labels={\strut $\infty$,\strut \xmark},%
  extra y ticks={3e2,700}, extra y tick labels={\strut $\infty$,\strut \xmark},%
  extra tick style={
      grid=major,
      line width=.2pt,draw=gray!50
  },
  xlabel near ticks,
  ylabel near ticks,
}}

\pgfplotsset{runtime diagram marks/.style={
  only marks,
  mark=x,
  mark size=2.5,
}}

\begin{figure}[th]
  \centering
  \scalebox{0.9}{
    \begin{tikzpicture}
      \begin{loglogaxis}[
        name=qdtvsspacer,
        anchor=north west,
        runtime diagram,
        height=45mm,
        xlabel = {\textsc{Spacer} (time in s)},
        ylabel = {\textsc{Tapis} (time in s)},
      ]
    
      \addplot[runtime diagram marks] table[col sep=comma, x={spacer}, y={tapis}] {results.csv};
    
      % Diagonal 
      \draw[black!25] (rel axis cs:0, 0) -- (rel axis cs:1, 1);
    
      \end{loglogaxis}

      \begin{loglogaxis}[
        name=qdtvsfreqhorn,
        at={(qdtvsspacer.north east)}, anchor=north west, xshift=30mm,
        runtime diagram,
        height=45mm,
        xlabel = {\textsc{Prophic3} (time in s)},
        ylabel = {\textsc{Tapis} (time in s)},
      ]
    
      \addplot[runtime diagram marks] table[col sep=comma, x={prophic3}, y={tapis}] {results.csv};
    
      % Diagonal 
      \draw[black!25] (rel axis cs:0, 0) -- (rel axis cs:1, 1);
    
      \end{loglogaxis}

      \begin{loglogaxis}[
        name=qdtvsvajra,
        at={(qdtvsspacer.south)}, anchor=north, yshift=-12mm,
        runtime diagram,
        height=45mm,
        xlabel = {\textsc{Vajra} (time in s)},
        ylabel = {\textsc{Tapis} (time in s)},
      ]
    
      \addplot[runtime diagram marks] table[col sep=comma, x={vajra}, y={tapis}] {results.csv};
    
      % Diagonal 
      \draw[black!25] (rel axis cs:0, 0) -- (rel axis cs:1, 1);
    
      \end{loglogaxis}

      \begin{loglogaxis}[
        at={(qdtvsvajra.north east)}, anchor=north west, xshift=30mm,
        runtime diagram,
        height=45mm,
        xlabel = {\textsc{Diffy} (time in s)},
        ylabel = {\textsc{Tapis} (time in s)},
      ]
    
      \addplot[runtime diagram marks] table[col sep=comma, x={diffy}, y={tapis}] {results.csv};
    
      % Diagonal 
      \draw[black!25] (rel axis cs:0, 0) -- (rel axis cs:1, 1);
    
      \end{loglogaxis}

      \begin{loglogaxis}[
        name=qdtvsua,
        at={(qdtvsvajra.south)}, anchor=north, yshift=-12mm,
        runtime diagram,
        height=45mm,
        xlabel = {\textsc{UAutomizer} (time in s)},
        ylabel = {\textsc{Tapis} (time in s)},
      ]
    
      \addplot[runtime diagram marks] table[col sep=comma, x={ultimate-automizer}, y={tapis}] {results.csv};
    
      % Diagonal 
      \draw[black!25] (rel axis cs:0, 0) -- (rel axis cs:1, 1);
    
      \end{loglogaxis}

      \begin{loglogaxis}[
        at={(qdtvsua.north east)}, anchor=north west, xshift=30mm,
        runtime diagram,
        height=45mm,
        xlabel = {\textsc{FreqHorn} (time in s)},
        ylabel = {\textsc{Tapis} (time in s)},
      ]
    
      \addplot[runtime diagram marks] table[col sep=comma, x={freqhorn}, y={tapis}] {results.csv};
    
      % Diagonal 
      \draw[black!25] (rel axis cs:0, 0) -- (rel axis cs:1, 1);
    
      \end{loglogaxis}
    \end{tikzpicture}
  }
\setlength{\abovecaptionskip}{0ex}
\setlength{\belowcaptionskip}{-3ex}
  \caption{Runtime of \textsc{Tapis} vs. \textsc{Spacer}, \textsc{Prophic3}, \textsc{Vajra}, \textsc{Diffy}, \textsc{UAutomizer} and \textsc{FreqHorn}.}
  \label{fig:runtime}
\end{figure}
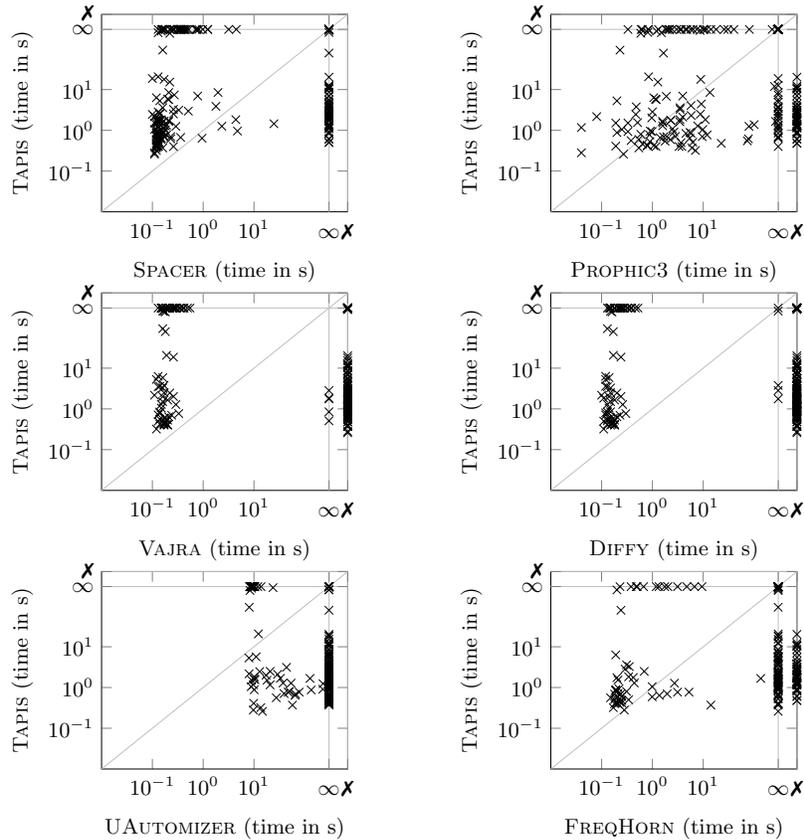

The plots in Fig.~\ref{fig:runtime} compare execution times of \textsc{Tapis} with \textsc{Spacer}, \textsc{Prophic3}, \textsc{Vajra}, \textsc{Diffy}, \textsc{UAutomizer} and \textsc{FreqHorn}.
Instances at~$\infty$ could not be solved by the tool in 300s (timeout) and \xmark~indicates instances that are not
in the class of programs verifiable by the tool.
They show that \textsc{Tapis} has comparable execution time with \textsc{Prophic3}, \textsc{FreqHorn} and \textsc{Diffy}, it is faster than \textsc{UAutomizer} and is slightly slower than \textsc{Spacer}, \textsc{Vajra} and \textsc{Diffy}.
Globally, the experiments show that \textsc{Tapis} is able to verify a large class of programs with competitive execution times compared to the state-of-the-art.
%The decision-tree construction from a large number of syntactically enumerated attributes is the main performance issue of \textsc{Tapis}.
%Additionally, its effectiveness is limited for unsafe programs, as it is not specifically designed for their verification.
%However, it shines in its capability to verify a large class of programs with reasonable execution times compared to the state-of-the-art.
%\iflong
%A detailed evaluation of our experiments is in Appendix~\ref{appendix:benchmarks}.
%\else
%A detailed evaluation is provided in Appendix~C of the extended version of this paper in the supplementary material.
%\fi

% 8. Conclusion
\section{Conclusion}
We have proposed an efficient data-driven method for the verification of programs with arrays based on a powerful procedure  for learning universally quantified loop invariants and procedure pre/post-conditions for array-manipulating programs, extending the Horn-ICE framework.
The experimental results are encouraging.
They show that our approach is efficient, solving globally more cases than existing tools on a significant benchmark, and that it is complementary to other approaches as it can deal with programs that could not be solved by state-of-the-art tools.
For future work, several issues need to be addressed such as improving the generation of relevant attributes in decision-tree learning and handling quantifier alternation.

% the bibliography file.
\bibliographystyle{splncs04} % LNCS: style
\bibliography{bibliography}

\iflong
  \newpage
  \appendix
  \section{Details of the Bubble Sort Example}
\label{appendix:bubble-sort-iterations}
The table below presents a detailed interaction between the teacher and the learner during the verification of the program from the overview (Section~\ref{sec:overview}).
For each iteration, it displays the candidate solution proposed by the learner and the counterexample provided by the teacher to refine the learner's hypothesis.

%\begin{adjustbox}{center}
\begin{table}
{
\tiny
\begin{tabular}{ m{0.3cm} || >{\RaggedRight}m{6.8cm} | >{\RaggedRight}m{4.6cm} }
\hline
 Iter. & Learner proposition & Teacher counterexample \\\hline\hline
 1 & \shortstack{$\Predicate{I_0}:\; \Interpretation{N,a,s}{\top}$ \\ $\Predicate{I_1}:\; \Interpretation{N,a,s,i}{\top}$} & $\langle \Predicate{I_0}, 2, [1, 0], \bot \rangle \rightarrow \bot$ \\\hline
 2 & \shortstack{$\Predicate{I_0}:\; \Interpretation{N,a,s}{\forall k_1,k_2.\; 0 \leq k_1 \leq k_2 < \BoundedArrayLength{a} \implies a[k_1] > 0}$ \\ $\Predicate{I_1}:\; \Interpretation{N,a,s,i}{\top}$} & $\top \rightarrow \langle \Predicate{I_0}, 1, [0], \top \rangle$ \\\hline
 3 & \shortstack{$\Predicate{I_0}:\; \Interpretation{N,a,s}{\forall k_1,k_2.\; 0 \leq k_1 \leq k_2 < \BoundedArrayLength{a} \implies k_1 \leq 0}$ \\ $\Predicate{I_1}:\; \Interpretation{N,a,s,i}{\top}$} & $\top \rightarrow \langle \Predicate{I_0}, 2, [0,0], \top \rangle$ \\\hline
 4 & \shortstack{$\Predicate{I_0}:\; \Interpretation{N,a,s}{\forall k_1,k_2.\; 0 \leq k_1 \leq k_2 < \BoundedArrayLength{a} \implies a[k_2] \leq 0}$ \\ $\Predicate{I_1}:\; \Interpretation{N,a,s,i}{\top}$} & $\top \rightarrow \langle \Predicate{I_0}, 1, [1], \top \rangle$ \\\hline
 5 & \shortstack{$\Predicate{I_0}:\; \Interpretation{N,a,s}{\forall k_1,k_2.\; 0 \leq k_1 \leq k_2 < \BoundedArrayLength{a} \implies a[k_2] \leq 0 \lor s}$ \\ $\Predicate{I_1}:\; \Interpretation{N,a,s,i}{\top}$} & $\langle \Predicate{I_1}, 1, [1], 1, \bot \rangle \rightarrow \langle \Predicate{I_0}, 1, [1], \bot \rangle$ \\\hline%
 6 & \shortstack{$\Predicate{I_0}:\; \Interpretation{N,a,s}{\forall k_1,k_2.\; 0 \leq k_1 \leq k_2 < \BoundedArrayLength{a} \implies k_1 \leq 0 \lor s}$ \\ $\Predicate{I_1}:\; \Interpretation{N,a,s,i}{\top}$} & $\langle \Predicate{I_1}, 2, [0,0], 2, \bot \rangle \rightarrow \langle \Predicate{I_0}, 2, [0,0], \bot \rangle$ \\\hline%
 7 & \shortstack{$\Predicate{I_0}:\; \Interpretation{N,a,s}{\forall k_1,k_2.\; 0 \leq k_1 \leq k_2 < \BoundedArrayLength{a} \implies a[k_1] \leq a[k_2]}$ \\ $\Predicate{I_1}:\; \Interpretation{N,a,s,i}{\top}$} & $\top \rightarrow \langle \Predicate{I_0}, 2, [1,0], \top \rangle$ \\\hline
 8 & \shortstack{$\Predicate{I_0}:\; \Interpretation{N,a,s}{\forall k_1,k_2.\; 0 \leq k_1 \leq k_2 < \BoundedArrayLength{a} \implies a[k_1] \leq a[k_2] \lor s}$ \\ $\Predicate{I_1}:\; \Interpretation{N,a,s,i}{\top}$} & $\langle \Predicate{I_1}, 2, [1,0], 2, \bot \rangle \rightarrow \langle \Predicate{I_0}, 2, [1,0], \bot \rangle$ \\\hline
 9 & \shortstack{$\Predicate{I_0}:\; \Interpretation{N,a,s}{\forall k_1,k_2.\; 0 \leq k_1 \leq k_2 < \BoundedArrayLength{a} \implies a[k_1] \leq a[k_2] \lor s}$ \\ $\Predicate{I_1}:\; \Interpretation{N,a,s,i}{\forall k_1,k_2.\; 0 \leq k_1 \leq k_2 < \BoundedArrayLength{a} \implies a[k_1] \leq a[k_2]}$} & $\langle \Predicate{I_0}, 2, [1,0], \top \rangle \rightarrow \langle \Predicate{I_1}, 2, [1,0], 1, \bot \rangle$ \\\hline
 10 & \shortstack{$\Predicate{I_0}:\; \Interpretation{N,a,s}{\forall k_1,k_2.\; 0 \leq k_1 \leq k_2 < \BoundedArrayLength{a} \implies a[k_1] \leq a[k_2] \lor s}$ \\ $\Predicate{I_1}:\; \Interpretation{N,a,s,i}{\forall k_1,k_2.\; 0 \leq k_1 \leq k_2 < \BoundedArrayLength{a} \implies a[k_1] \leq a[k_2] \lor i \leq k_2}$} & $\langle \Predicate{I_1}, 3, [1,1,0], 2, \bot \rangle \rightarrow \langle \Predicate{I_1}, 3, [1,0,1], 3, \top \rangle$ \\\hline
 11 & \shortstack{$\Predicate{I_0}:\; \Interpretation{N,a,s}{\forall k_1,k_2.\; 0 \leq k_1 \leq k_2 < \BoundedArrayLength{a} \implies a[k_1] \leq a[k_2] \lor s}$ \\ $\Predicate{I_1}:\; \Interpretation{N,a,s,i}{\forall k_1,k_2.\; 0 \leq k_1 \leq k_2 < \BoundedArrayLength{a} \implies a[k_1] \leq a[k_2] \lor i \leq k_2 \lor s}$} & Inductive invariants \\
 \hline
\end{tabular}
}
\end{table}

\section{Example of recursive program}
\label{appendix:rec-program}
Fig.~\ref{fig:max} shows an example of a program with its specification that has a recursive procedure.
It computes the index of a maximal element of an array of parametric size $N$ using a recursive procedure.
%Notice that formula $\forall k.\; 0 \leq k < N \implies a[k] \leq a[im]$ can be written equivalently as the parametric-size array property $\BoundedArrayLength{a} = N \wedge \forall k,k'.\; 0 \leq k,k' < \BoundedArrayLength{a} \implies (k' = im \implies a[k] \leq a[k'])$ where $N$ corresponds to the size of $a$.
  \begin{figure}[ht]
    \centering
    \scriptsize
    \begin{minipage}{0.45\textwidth}
      \begin{lstlisting}
void main() {
  int N;
  assume(N > 0);
  int  a[N];
  int im = argmax(a, 0, N);
  assert($\forall k.\; 0 \leq k < N \implies a[k] \leq a[im]$);
}
      \end{lstlisting}
    \end{minipage}
    \quad
    \begin{minipage}{0.47\textwidth}
      \begin{lstlisting}
int argmax(int a[], int i, int N) {
  if(i == N - 1) {
    return i;
  }
  int im = argmax(a, i + 1, N);
  if(a[i] $\geq$ a[im]) {
    return i;
  }
  return im;
}
      \end{lstlisting}
    \end{minipage}
    \caption{A program that computes the index of the maximal element of an array (left) using a recursive procedure (right).}
    \label{fig:max}
  \end{figure}

Our method verifies the safety of this program and finds a precondition $\Predicate{P}$ and a postcondition $\Predicate{Q}$ for the procedure \texttt{argmax}:
\begin{align*}
\Predicate{P}: & \; 0 \leq i < N \land \BoundedArrayLength{a} = N \\
\Predicate{Q}: & \; 0 \leq i < N \land \forall k. i \leq k < N \implies \BoundedArrayRead{a}{k} \leq \BoundedArrayRead{a}{res}
\end{align*}

\section{Proofs}
\label{appendix:proofs}
\setcounter{theorem}{0}

\newcommand{\ICEImplication}[2]{{#1} \rightarrow {#2}}

\begin{theorem}
  Let $\mathcal{S} = (X, C)$ be a consistent data point sample and let
  $\mathcal{S}'_n = \delta_n(\mathcal{S})$ be the corresponding diagram sample.
  If $\mathcal{S}'_n$ is consistent and $J'$ is a classifier of $\mathcal{S}'_n$ then~$\xi(J')$ is a classifier of $\mathcal{S}$.
\end{theorem}
\begin{proof}
  Let $\mathcal{S}'_n = \delta_n(\mathcal{S}) =  (X', C')$ the corresponding consistent diagram sample obtained from the consistent data point sample~$\mathcal{S} = (X, C)$, and let $J'$~be a classifier of~$\mathcal{S}'_n$.\\
  To prove that $J = \xi(J')$~is a classifier of~$\mathcal{S}$, we must show that~$J$ satisfies every implication in~$C$.\\
  Let~$\mathcal{J}$ (respectively,~$\mathcal{J}'$) be the consistent labeling that~$J$ (respectively,~$J'$) syntactically characterizes.\\
  For every implication~$c \in C$, we have three separate cases depending on the form of~$c$.
  \begin{itemize}
    % Case 1: \top -> x
    \item {\bf (Case~1)} Let $c$ be of the form $\top \rightarrow x$ and~$\Predicate{P}$ is the predicate of~$x$.
    We must show that $x \models \xi(J')[\Predicate{P}]$. \\
    We prove this by contradiction.
    Assume  the opposite holds: $x \not\models \xi(J')[\Predicate{P}]$.
    By the definition of~$\xi(J')$, this means that
    \begin{equation*}
      x \models \exists \vec{Q}_{a_1}, \ldots ,\vec{Q}_{a_n} .\; \big(\bigwedge_{i=1}^n \bigwedge_{\;\;k \in \vec{Q}_{a_i}} 0 \leq k < \BoundedArrayLength{x[a_i]} \big) \land \neg J'[\Predicate{P}][a_k/x[a][k],l_{a}/|x[a]|]_{a \in \ArraysOf{\Predicate{P}}}
    \end{equation*}
    Hence there is a model~$m$ that assigns to every quantifier variable~$k \in \vec{Q}_{a_1} \cup \cdots \cup \vec{Q}_{a_n}$ a value in the range~$\Interval{0}{|x[a_j]|-1}$ (for $k \in \vec{Q}_{a_j}$) and $m$~falsifies $J'[\Predicate{P}]$.\\
    By the definition of diagrams (Definition~\ref{def:diagram}), there is a diagram~$d \in \Diagrams{n}{x}$ whose quantifer variables and scalar values match the assignments in~$m$.
    Since $m$~falsifies $J'[\Predicate{P}]$ then $d \not\models J'[\Predicate{P}]$ ($\mathcal{J}'(d) = \bot$). \\
    However, since~$\top \rightarrow x$ is in~$C$, then by the construction of~$\mathcal{S}'_n$, there exists an implication~$\top \rightarrow d$ in~$C'$. $\mathcal{J}'(d) = \bot$ is contradicting the hypothesis that $J'$~is a classifier of~$\mathcal{S}'_n$ becauase $J'$ is a syntactic characterization of~$\mathcal{J}'$.
    Therefore, $x \models \xi(J')[\Predicate{P}]$ and $\xi(J')$ satisifes~$c$.

    % Case 2: x1,...,xm -> \bot
    \item {\bf Case~2} Let $c$ be of the form $x_1 \land \dots \land x_m \rightarrow \bot$.
    For each $x_j$, let $\Predicate{P_j}$ denote its predicate.
    We must show that $x_1 \not\models \xi(J')[\PredicateOfDatapoint{x_1}] \lor \cdots \lor x_m \not\models \xi(J')[\PredicateOfDatapoint{x_m}]$. \\
    We prove this by contradiction.
    Assume  the opposite holds: $x_1 \models \xi(J')[\PredicateOfDatapoint{x_1}] \land \cdots \land x_m \models \xi(J')[\PredicateOfDatapoint{x_m}]$. \\
    As in Case~1, we conclude that for every $d \in \Diagrams{n}{x_j}$, $\mathcal{J}'(d) = \top$.
    Since~$c \in C$, there is, by the construction of~$\mathcal{S}'_n$, in~$C'$ the implication;
    \begin{equation*}
      \bigwedge_{d \in \bigcup_{i=1}^m \Diagrams{n}{x_i}} d \rightarrow \bot
    \end{equation*}
    For this implication to be satisfied by~$J'$, there must exists a diagram~$d \in \bigcup_{i=1}^m \Diagrams{n}{x_i}$ such that~$d \not\models J'[\PredicateOfDatapoint{d}]$ ($\mathcal{J}'(d) = \bot$) and let $x_l$ the data points such that $d \in \Diagrams{n}{x_l}$.
    This contradicts the previous conclusion stating that for every $d \in \Diagrams{n}{x_j}$, $\mathcal{J}'(d) = \top$.\\
    Therefore, our assumption is false and consequently $x_1 \not\models \xi(J')[\PredicateOfDatapoint{x_1}] \lor \cdots \lor x_m \not\models \xi(J')[\PredicateOfDatapoint{x_m}]$ and $\xi(J')$ satisifes~$c$.

    % Case 3: x1,...,xn -> xm
    \item {\bf (Case~3)} Let $c$ be of the form $x_1 \land \dots \land x_m \rightarrow x_j$.
    For each $x_j$, let $\Predicate{P_j}$ denote its predicate.
    There are two sub-scenarios:
    \begin{itemize}
      \item If $\mathcal{J}(x_1) = \top \land \cdots \land \mathcal{J}(x_m) = \top$, then we follow as in Case~1 to show that also $\mathcal{J}(x_j) = \top$.

      \item Conversely, if $\mathcal{J}(x_j) = \bot$, then as in Case~2, we conclude that $\mathcal{J}(x_1) = \bot \lor \cdots \lor \mathcal{J}(x_m) = \bot$.
    \end{itemize}
    In the both sub-scenarios, $\xi(J')$ satisifes~$c$.
  \end{itemize}
  We conclude, finally, that $J = \xi(J')$ satisfies every implication of~$C$.
  Hence $J$~is a classifier to~$\mathcal{S}$.
\end{proof}

\begin{definition}[Complete Diagram]
  Let~$x$ be a data point whose predicate is~$P$, and assume that for each array~$a \in \ArraysOf{\Predicate{P}}$, we have~$|x[a]| \leq |\ArrayQuantifierSet{a}{\Predicate{P}}|$.\\
  A \emph{complete diagram} $d$ of $x$ is any diagram in $\Diagrams{n}{x}$ with the following property:
  \begin{itemize}
    \item For every array $a \in \ArraysOf{\Predicate{P}}$, we pick a mapping
    \begin{equation*}
      f_a \colon \Interval{0}{|x[a]|-1} \to \ArrayQuantifierSet{a}{\Predicate{P}}
    \end{equation*}
    that assigns each integer $i \in \Interval{0}{|x[a]|-1}$ to a distinct quantifier variable $k = f_a(i)$.\\
    Then $d[k] = i$ and $d[a_k] = x[a][i]$. \\
    In addition $d[l_a] = |x[a]|$ for the size variable of~$x$.

    \item For every non-array variable $v \in \VariablesOf{\Predicate{P}} \setminus \ArraysOf{\Predicate{P}}$, $d[v] = x[v]$.
  \end{itemize}
\end{definition}
\begin{example}
  $\langle \Predicate{I_0}, N \mapsto 2, s \mapsto \bot, k_1 \mapsto 0, k_2 \mapsto 1, a_{k_1} \mapsto 1, a_{k_2} \mapsto 0, l_{a} \mapsto 2 \rangle$ is the complete diagram of $\langle \Predicate{I_0}, N \mapsto 2, a \mapsto [1,0], s \mapsto \bot \rangle$.
\end{example}

\begin{lemma} \label{lemma:complete-diagram-uniqueness}
  If $x$ and $y$ are two data points of the same predicate~$\Predicate{P}$ such that $x \neq y$, then no complete diagram of~$x$ can be the same as a complete diagram of~$y$.
\end{lemma}
\begin{proof}
  We prove this by contradiction, suppose that $d$~is complete diagram of both~$x$ and~$y$ such that~$x \neq y$.\\
  Since~$x \neq y$, either:
  \begin{enumerate}
    \item They differ in a non-array variable.
    Let $v$~be a variable in~$\VariablesOf{\Predicate{P}} \setminus \ArraysOf{\Predicate{P}}$ for which $x[v] \neq y[v]$.\\
    By the definition of a diagram
    \begin{equation*}
      d[v] = x[v] \text{ and } d[v] = y[v]
    \end{equation*}
    This implies that~$x[v] = y[v]$ which contradicts the assumption that~$x[v] \neq y[v]$

    \item They differ in an array size.
    Let $a$~be an array in~$\ArraysOf{\Predicate{P}}$ for which $|x[a]| \neq |y[a]|$.\\
    By the definition of a diagram
    \begin{equation*}
      d[l_a] = |x[a]| \text{ and } d[l_a] = |y[a]|
    \end{equation*}
    This implies that~$|x[v]| = |y[v]|$ which contradicts the assumption that~$|x[a]| \neq |y[a]|$.

    \item They differ in an array value.
    Let $a$~be an array in~$\ArraysOf{\Predicate{P}}$ for which $|x[a]| = |y[a]|$ and there exists some~$i$ such that~$i \in \Interval{0}{|a|} \land x[a][i] \neq y[a][i]$.\\
    Since $d$ is a complete diagram of~$x$ then there exists some quantifier variable $k \in \ArrayQuantifierSet{a}{\Predicate{P}}$ such that
    \begin{equation*}
      d[k] = i \text{ and } d[a_k] = x[a][i]
    \end{equation*}
    But since $d$ is also a complete diagram of~$y$ then $d[k] = i$ and $d[a_k] = y[a][i]$ wich contradicts the assumption that $x[a][i] \neq y[a][i]$.
  \end{enumerate}
  In all scenarios, we reach a contradiction.
  Therefore, a diagram~$d$ cannot be a complete diagram for both $x$ and $y$ whenever~$x \neq y$.
\end{proof}

\begin{theorem}
  Let~$\mathcal{S} = (X, C)$ be a consistent data point sample over $\mathcal{P}$.
  Suppose that for every predicate $P \in \mathcal{P}$ and for every array~$a$ in the domain of~$P$, we have~$n \geq \max_{x \in X} |x[a]|$.
  Then the diagram sample~$\delta_n(\mathcal{S})$ is also consistent. Equivalently, there exists a consistent labeling~$\mathcal{J}'$ for~$\delta_n(\mathcal{S})$.
\end{theorem}
\begin{proof}
  Since the data point sample~$\mathcal{S} = (X, C)$ is consistent, it admits a consistent labeling~$\mathcal{J}$ that satisfies all the implications in~$C$.\\
  We want to construct a consistent labeling~$\mathcal{J}'$ for~$\mathcal{S}'_n = \delta_n(\mathcal{S}) = (X', C')$ under the assumption that each array~$a$ of a predicate $\Predicate{P}$ has at least as many quantifier variables as the length of~$a$ in each data point~$x \in X$ of~$\Predicate{P}$:
  \begin{equation*}
    n \geq |\ArrayQuantifierSet{a}{\Predicate{P}}| \geq |x[a]|
  \end{equation*}
  as follow and in two steps:
  \begin{enumerate}
    \item For every data point~$x \in X$ such that $\mathcal{J}(x) = \top$, we assign $\mathcal{J}'(d) = \top$ for every diagram~$d \in \Diagrams{n}{x}$.

    \item For every data point~$x \in X$ such that $\mathcal{J}(x) = \bot$, some of the diagrams of $x$ may already have been assigned $\top$ in $\mathcal{J}'$ by the previous step (multiple data points may share a same diagram).
    We assign $\mathcal{J}'(d) = \bot$ for every remaining diagrams of~$d \in \Diagrams{n}{x}$, the remaining diagrams include at least the complete diagrams of $x$, which surely have not been assigned in previous step (as justified with the Lemma~\ref{lemma:complete-diagram-uniqueness}).
  \end{enumerate}

  We now prove that the constructed~$\mathcal{J}'$ is a classifier for~$\mathcal{S}'_n = \delta_n(\mathcal{S})$ by showing that $\mathcal{J}'$~satisifies all the implications in~$C$.
  Recall that each implication in~$C'$ arises from some implication in $C$.
  For each implication~$c \in C'$:
  \begin{itemize}
    \item {\bf (Case~1)} If $c$~is of the form $\top \rightarrow d$, it is constructed from an implication $\top \rightarrow x$ in~$C$ and $d \in \Diagrams{n}{x}$. \\
    Since $\mathcal{J}$~is a consistent labeling for~$\mathcal{S}$ then $\mathcal{J}(x) = \top$, and following the construction then $\mathcal{J'}(d) = \top$.
    Hence $\mathcal{J}'$~satisfies~$c$.

    \item {\bf (Case~2)} If $c$~is of the form $d_1 \land \dots \land d_m \rightarrow \bot$, it is constructed from an implication $x_1 \land \dots \land x_l \rightarrow \bot$ in~$C$ such that $\{d_1,\ldots,d_m\} = \bigcup_{x \in \{x_1,\ldots,c_l\}} \Diagrams{n}{x}$. \\
    Since $\mathcal{J}$~is a consistent labeling for~$\mathcal{S}$ then there exists some $x_j \in \{x_1,\ldots,c_l\}$ such that $\mathcal{J}(x_j) = \bot$, and following the construction then there exists a complete diagram~$d \in \Diagrams{n}{x_j}$ such that $\mathcal{J'}(d) = \bot$, $d \in \{d_1,\ldots,d_m\}$ since $\Diagrams{n}{x_j} \subseteq \{d_1,\ldots,d_m\}$.
    Hence $\mathcal{J}'$~satisfies~$c$ (at least one of the diagrams in $d_1,\ldots,d_m$ is classifed as false in $\mathcal{J}'$).

    \item {\bf (Case~3)} If $c$~is of the form $d_1 \land \dots \land d_m \rightarrow d$, it is constructed from an implication $x_1 \land \dots \land x_l \rightarrow x$ in~$C$ such that $\{d_1,\ldots,d_m\} = \bigcup_{x \in \{x_1,\ldots,c_l\}} \Diagrams{n}{x}$ and $d \in \Diagrams{n}{x}$. \\
    If for all $x_i \in \{x_1,\ldots,x_l\}$, $\mathcal{J}(x_i) = \top$, we follow the same reasoning as in {\bf (Case~1)} to prove that $\mathcal{J}'(d) = \top$.
    Hence $\mathcal{J}'$~satisfies~$c$. \\
    Otherwise if $\mathcal{J}'(d) = \bot$, we follow the same reasoning as in {\bf (Case~2)} to prove that for at least for a diagram $d_i \in \{d_1,\ldots,d_m\}$, $\mathcal{J}'(d_i) = \bot$.
    Hence $\mathcal{J}'$~satisfies~$c$. \\
  \end{itemize}
  In all cases, we find out that $\mathcal{J}'$~satisfies all the implications in~$C'$.
  Therefore, $\mathcal{S}'_n = \delta_n(\mathcal{S})$~is consistent and admits a consistent labeling $\mathcal{J}'$.
\end{proof}

\else
\fi

\end{document}